%% file: main.tex
\documentclass[12pt,a4paper,openright]{book}
\usepackage{graphicx}
\usepackage{background}
\usepackage{fancyhdr}   
\usepackage{multicol}
\usepackage{makeidx}
\usepackage{hyperref} 
\usepackage{amsmath}
\usepackage{amssymb}
\usepackage{pstricks,pst-node}
\usepackage{subfigure}
\usepackage{setspace}
\usepackage{lineno}
\usepackage{cite}
\usepackage{subfigure}
\usepackage{latexsym}
\usepackage{setspace}
\usepackage[top=25mm, left=35mm, bottom=25mm, right=25mm]{geometry}
\usepackage{slashbox}
\usepackage{caption}
\usepackage{rotating}
\usepackage{enumerate}
\usepackage{afterpage}
\usepackage{mwe}
\usepackage{xcolor}
\usepackage{colortbl}
\usepackage{multirow}
\usepackage{mathtools}
\usepackage[b]{esvect}
\usepackage{gensymb}

\usepackage{authblk}
\usepackage{lipsum}

\usepackage{environ}

\NewEnviron{NORMAL}{%
    \scalebox{2}{$\BODY$}
}
 
\NewEnviron{HUGE}{%
    \scalebox{5}{$\BODY$}
}

\newcounter{relctr} 
\everydisplay\expandafter{\the\everydisplay\setcounter{relctr}{0}} 

\AtBeginDocument{} 

\makeatletter
\newsavebox\myboxA
\newsavebox\myboxB
\newlength\mylenA

\newcommand*\xoverline[2][0.75]{%
    \sbox{\myboxA}{$\m@th#2$}%
    \setbox\myboxB\null
    \ht\myboxB=\ht\myboxA%
    \dp\myboxB=\dp\myboxA%
    \wd\myboxB=#1\wd\myboxA
    \sbox\myboxB{$\m@th\overline{\copy\myboxB}$}
    \setlength\mylenA{\the\wd\myboxA}
    \addtolength\mylenA{-\the\wd\myboxB}%
    \ifdim\wd\myboxB<\wd\myboxA%
       \rlap{\hskip 0.5\mylenA\usebox\myboxB}{\usebox\myboxA}%
    \else
        \hskip -0.5\mylenA\rlap{\usebox\myboxA}{\hskip 0.5\mylenA\usebox\myboxB}%
    \fi}
\makeatother

\def\dept{Department of Electrical Engineering}
\def\IIT{Indian Institute of Technology Kanpur}
\def\instaddress{Kanpur, INDIA - 208016}
\title{Towards Fast, Flexible and Sensor-Free Control of Standalone PVDG Systems}
\author{MEHER PREETAM KORUKONDA}
\def\rollno{12104172}
\def\thesisdate{June, 2020}                       
\def\supervisor{Laxmidhar Behera}
\def\supervisorPosition{Professor}

\backgroundsetup{contents={}}

\makeindex

\begin{document}
\include{Content/titlepage}
\doublespacing
\onehalfspacing
\newpage \null\thispagestyle{empty}  
\include{Content/certificate}
\include{Content/Declarationbest}
\include{Content/abstract}
\let\cleardoublepage\clearpage
\include{Content/acknowledgements}
\tableofcontents
\listoffigures \addcontentsline{toc}{chapter}{List of Figures}
\listoftables \addcontentsline{toc}{chapter}{List of Tables}

\mainmatter
\pagestyle{headings}
\pagenumbering{arabic}

\include{Content/01}  			
\include{Content/02}
\include{Content/03}

\include{Content/04}   		

\include{Content/05}
\include{Content/06}

\include{Content/Publications}

\addcontentsline{toc}{chapter}{Bibliography}
\bibliographystyle{IEEEbib}
\bibliography{refs}
\clearpage
\phantomsection
\addcontentsline{toc}{chapter}{Index}
\printindex

\end{document}

%% file: Content/titlepage.tex
\makeatletter
\thispagestyle{empty}
\begin{titlepage}
\begin{center}
\LARGE
\textbf{Towards Fast, Flexible and Sensor-Free Control of Standalone PVDG Systems} \\
\vspace{2cm}
\Large
{\em A Thesis Submitted}\\ in Partial Fulfillment of the Requirements\\
for the Degree of\\ \textbf{Master of Technology}\\
\vspace{2cm}
by\\
\textbf{Meher Preetam Korukonda} \\\textbf{(12104172)}
\vfill
\begin{figure}[!h]
\centering
\includegraphics[width=0.35\linewidth]{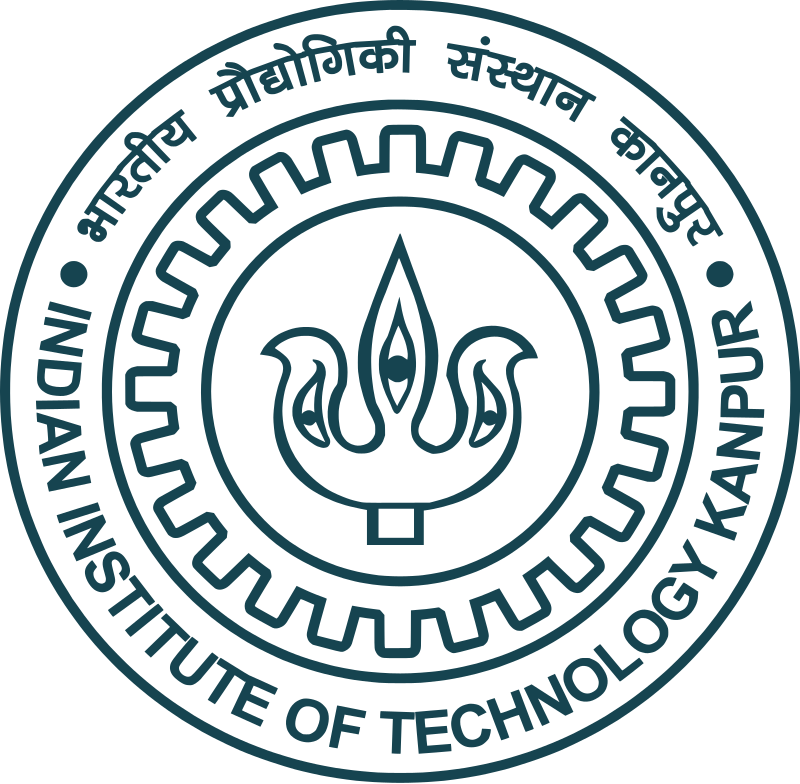}
\label{fig:spvdg0}
\end{figure}
{\em to the} \\
\dept \\
\MakeUppercase{\IIT} \\
\instaddress\\
\thesisdate 

\end{center}
\end{titlepage}
\makeatother

%% file: Content/Declarationbest.tex
\vspace {1.0in}
\begin{center}
\begin{large}
{\bf \underline{DECLARATION}}
\end{large}
\end{center}
\vskip 0.2in
 This is to certify that the thesis titled {\bf{\textit{Towards Fast, Flexible and Sensor-Free Control of Standalone PVDG Systems}}} has been authored by me. It presents the research conducted by me under the supervision of Prof. Laxmidhar Behera.\\
To the best of my knowledge, it is an original work, both in terms of research content and narrative, and has not been submitted elsewhere, in part or in full, for a degree. Further, due credit has been attributed to the relevant state-of-the-art and collaborations (if any) with appropriate citations and acknowledgments, in line with established norms and practices. \\
\pagestyle{plain}
\vspace{1cm}
\begin{minipage}{1.5\textwidth}
\vskip 2.0in
\begin{flushright}
\centering
\includegraphics[width=0.15\linewidth]{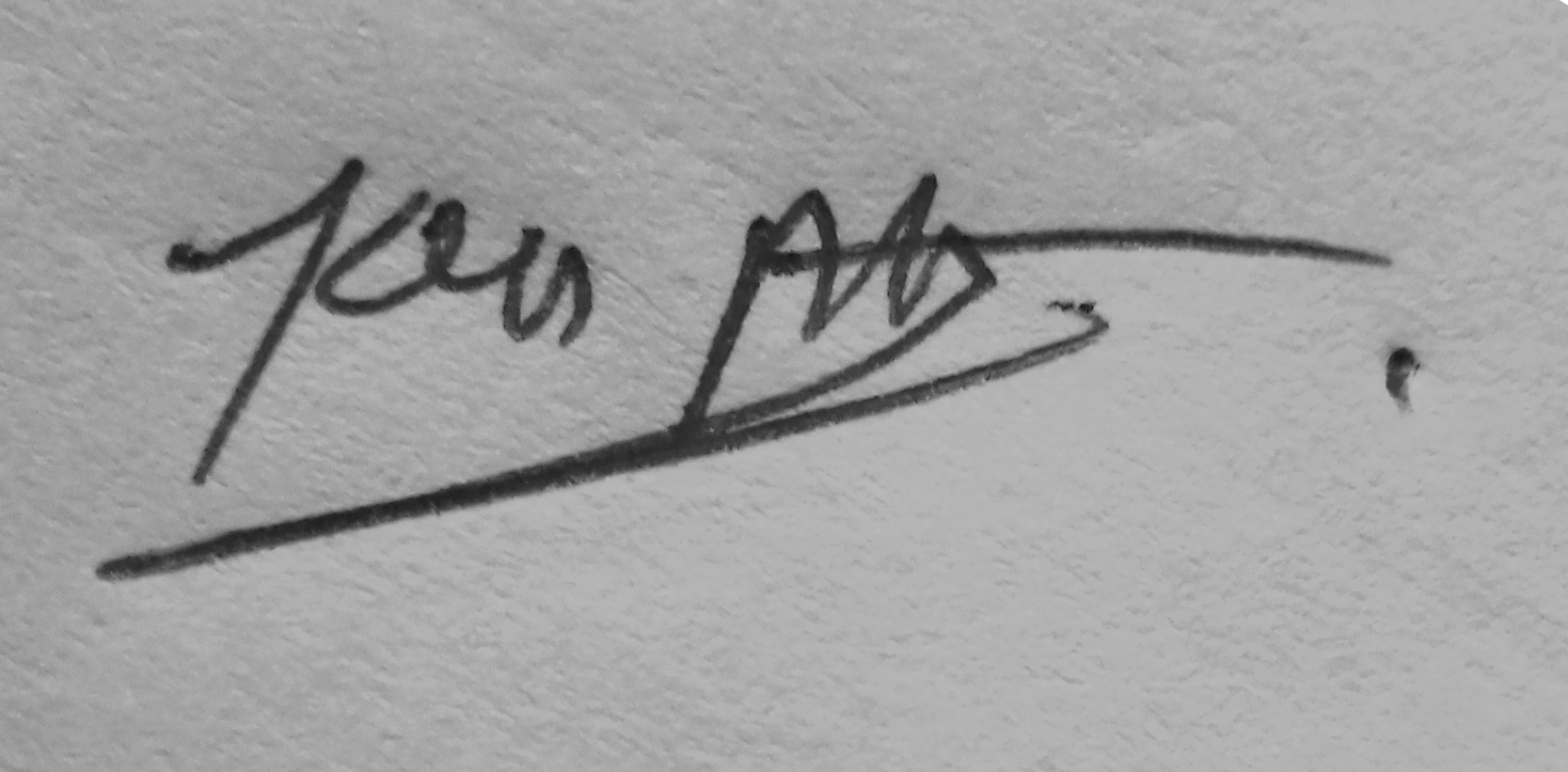}\\
	\vskip 0.2in
             Signature \hspace*{0cm} \\
		Name : Meher Preetam Korukonda \hspace*{0cm} \\
		Programme: MTech \hspace*{0cm} \\
		Department: Electrical Engineering \hspace*{0.01cm} \\
             Indian Institute of Technology Kanpur,\\
             Kanpur-208016, India \hspace*{0cm}
\end{flushright}
\end{minipage}

%% file: Content/abstract.tex
\emph{•}
\begin{center}
\begin{large}
{\it{\bf ABSTRACT} }
\end{large}
\end{center} 
\hrule
\vspace{1ex}
\noindent
Name of Student: Meher Preetam Korukonda \hspace{5ex}Roll no.: 12104172\\
Degree for which submitted: M.Tech\hspace{9ex}Department: Electrical Engineering\\
Title: Towards Fast, Flexible and Sensor-Free Control of Standalone PVDG Systems.\\
Name of Thesis Supervisor: Prof. Laxmidhar Behera

\hspace{-0.6cm}Month and year of Thesis submission: May, 2020.
\vspace{1ex}
\hrule

\vspace{3ex}
\noindent
In this thesis, the problem of fast, effective and low cost control of a Standalone Photovoltaic Distributed Generation (SPVDG) system is considered . On-site generation from these systems is more efficient when the power is transmitted via DC due to elimination of transmission losses and needless energy conversions. The inherent low-inertia of these systems added with fluctuation of output power and uncertain load consumption, calls for advanced control techniques to ensure fast and stable operation during various intermittencies. These techniques are expensive since they demand installation of many sophisticated sensors. The computation power provided by the fast growing IC technology can be utilized to estimate different parameters in a system and reduce the need for expensive sensing equipment. This work provides solutions to problems encountered in the development of faster, more stable and sensor-free voltage control and maximum power point tracking(MPPT) for SPVDG systems with PV and battery.

First, a model based MPPT technique fitted with a Newton-Raphson based temperature and irradiation estimation scheme is proposed for faster tracking and higher PV power extraction. Next, a unified direct perturbation based control algorithm is proposed which carries out both MPPT and DC bus voltage control simultaneously using voltage measurements only from the load side. Although this technique cuts down the overall cost, it suffers from poor performance in wake of large disturbances. To improve the speed and operating range of the SPVDG, a nonlinear back-stepping based control strategy is proposed which tackles various intermittencies in weather, load and battery voltage. However, this controller again suffers from higher cost as many parameters of the DCMG unit need to be explicitly measured. Hence, in the final part of the thesis, the back-stepping based controller design is revisited with inclusion of disturbance observers for grid voltage control. These observers estimate unknown parameters like load and battery voltage and eliminate additional measurements. It can be seen that adoption of the techniques proposed in this thesis contributes towards faster, sensor-free control of the DCMG for a greater range of operating conditions.

%% file: Content/acknowledgements.tex
\chapter*{\centering Acknowledgements}
\addcontentsline{toc}{chapter}{Acknowledgements}

Firstly, I would like to express my heartfelt gratitude to my guide, Prof. L. Behera for his valuable guidance and support throughout my work. He has provided with immense opportunities to explore and collaborate which finally took the form of this thesis. Apart from being a visionary, he has inspired me with the right mix of strictness and compassion to ensure my development in all walks of life. 

I wish to express my gratitude to the Bhaktivedanta Club where I was introduced teachings of H.D.G. A. C. Bhaktivedanta Swami Prabhupada and my spiritual master. I am highly obliged to all the devotees, in particular Mrs. Gopali, Dr.Vipul, Dr.V.Sudheendhra, Dr.Akhaya Nayak, Dr.Himanshu , Mr.Ashish Gupta, Dr. Jayant and Mr. Devaki Nandan  for guiding me through practical issues in life. I am especially grateful to Mr.Tharun Reddy, Mr.Ravi Prakash, Dr.Nitesh, Vinay Gupta, Jitendra Soni, Sachin Sahoo, Praful, Hariom and Jivnesh who readily took up many responsibilities to help me focus on my research. I wish to express my heartfelt thanks to Mr. Akshay Samal, Amruta, Sunil Dutta, Bharti, Sandeep Gupta, Prem Raj, Mrs. Swati Gupta,  Mrs.Sujata Samal and Mrs. Chitra whose loving association inspires me to be a better person.

I would like to express my deep gratitude to the esteemed faculty members of IIT Kanpur, who ignited within me, a deep sense of appreciation for engineering. Especially, I would like to thank Prof. P.Sensarma, Prof. Santanu Mishra, Prof. S.C.Srivastav, Prof. S.P. Das, and Prof. R. Potluri who taught me the basics of power electronics, power systems and control. 

I am highly grateful to our lab in-charges Abhay ji, Uday ji, Kamlesh ji, and assistant
 Harishankarji for their support.
I wish to thank Dr.Manmohan, Mr. Amir,Mr.Swaroop Mishra, Mr. Manoranjan, Dr.Narendra Dhar and Mr.Shamim for the stimulating discussions on microgrids. I would specially like to thank all my labmates: Ashish, Archit, Anuj, Vibhu, Padmini, Radheshyam, Subhash and especially my seniors Dr.Prem, Dr.Felix, Dr.Awhan, Dr. Samrat, Dr. Ranjith and Dr. Anima for extending me all help whenever I needed it.

Most importantly, I would like to thank my parents, and my brother for their placing their faith in me, encouraging me in challenging times while tolerating all the inconveniences I have put through them with love and patience.

%% file: Content/01.tex
\chapter{Introduction}\label{ch:introduction}
\pagenumbering{arabic}


\section{Motivation}
Renewable energy is penetrating the power generation sector today like never before. Especially, solar PVDG systems are gaining popularity due to a continuous reduction in the cost of solar panel,  efficiency improvement, advancement in power electronics, and, ambitious goals set by different countries to deploy PV sources into the existing electrical network  \cite{overview0}.Certain PVDG configurations, like the ones used in solar home projects in Africa (e.g., Kenya) \cite{kenyaref}, are seen to provide high-quality power to remote areas in a convenient and practical manner due to which many inaccessible areas gained access to cheap and reliable electricity supply.

PVDG systems can be categorized into two types, a) stand-alone system and b) grid connected system. The standalone PVDG system (SPVDG) \cite{overview2} is used in two different configurations, namely, with and without storage. In the SPVDG system without storage, the extracted power is directly supplied to the load. Incase of an SPVDG system with storage, the extracted power from PV is used both for feeding the load and charging the battery \cite{ramanuj} as well. An overview of the SPVDG system structure shown in Fig. \ref{fig:spvdg1}.  

Of late, deployment of DC based SPVDG systems and microgrids has gone upward since PVDG systems and many other renewable sources directly generate DC output power which can directly be connected to DC loads\cite{overviewtkr2}. If loads are supplied directly with DC power, the efficiency of the system becomes higher due to the reduction of conversion losses from sources to loads. Apart from these, DC implementation of SPVDG systems can also overcome some limitations associated with AC such as frequency synchronization, control of reactive power flow, and power quality problems \cite{overview6}. However, the characteristics of PV arrays are highly intermittent and so are the loading conditions in an SPVDG system and this makes control of these systems quite difficult \cite{overviewtkr1}. Storage can be used to eliminate the fluctuations in microgrids by storing or releasing energy \cite{overview4} using bidirectional DC–DC converters along with appropriate tuning of control parameters.

For a standalone PV system with storage, maximum power point tracking and DC bus voltage regulation constitute the major control problems to be dealt with. While MPPT ensures that maximum power is extracted from the PV array at any give time, DC bus voltage control indicates the overall power balance in the system. Research in the field of SPVDG has diversified into many areas such as  power electonic converter architecture \cite{flexarc1}, maximum-power-point tracking (MPPT) \cite{self1}, battery life expansion \cite{maheshess1}, efficiency improvement\cite{lhc}, power flow management of microgrids \cite{powermangment}, communication design\cite{self2} and distributed control\cite{self3,selfself1,selfself2,selfself3,selfself4}On the control side, many controllers have been designed for these systems including fractional PID \cite{fracpid}, and phase angle control\cite{varyphaseangle} apart from classical control techniques.

It has been noticed from all the above literature that model based nonlinear control techniques work much better when we need robust performance in MPPT and voltage control over large range of operating conditions. However, these techniques require many sophisticated sensors compared to conventional techniques which naturally increase the overall cost of the system. Hence, in this thesis, we use many computation based techniques to overcome some of the issues in this direction and provide better control solutions for the SPVDG system.

\section{System Overview} \index{SPVDG!Standalone Photovoltaic Distributed Generation}
The SPVDG system under consideration consists of a PV array and a battery energy storage system (BESS) feeding to a load. The PV array continuously extracts power from the available solar energy and transfers it to the load using a DC-DC converter as shown in Fig. \ref{fig:spvdg2}. The BESS is connected to the grid via a bidirectional DC-DC converter. This is essential to modulate power flow in both the directions as demanded by load conditions in the SPVDG system. Both the converters are controlled to regulate the desired flow of power in the PVDG system. 

\begin{figure}[!h]
\centering
\includegraphics[width=0.8\linewidth]{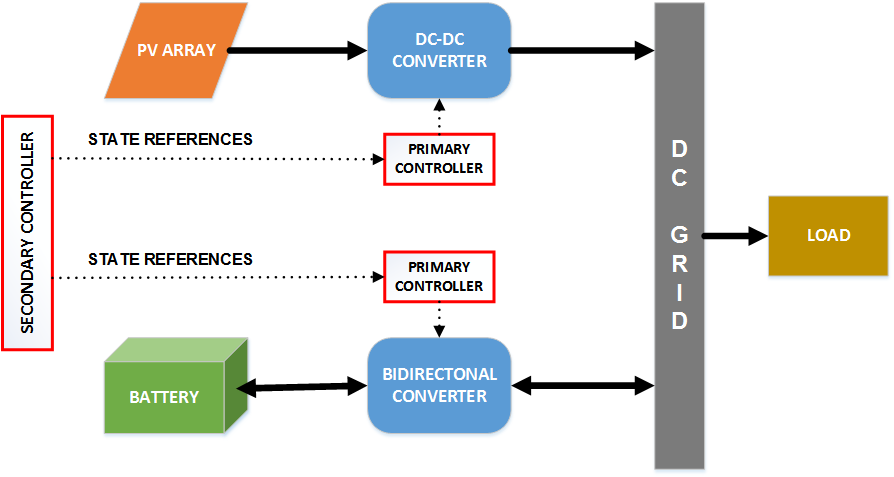}
\caption{Structure of SPVDG System}
\label{fig:spvdg00}
\end{figure}

 As evident from Fig.\ref{fig:spvdg00}, the hierarchy of control in the SPVDG system is present in two levels of the SPVDG system  namely, primary and secondary. The secondary level controller handles the overall energy management function of the SPVDG. It is responsible for ensuring maximum power extraction from the PV panel and also plan the power flow of the battery depending on the imbalance between the PV power generation and load consumption. The secondary controller achieves this operation by setting the reference values for all the important states in the SPVDG system like inductor currents and capacitor voltages. The major function of the primary controller is to bring the system states to the reference values set by the secondary controller. A properly designed primary controller performs this function even in the presence of large disturbances affecting the system time to time.

\section{Thesis Organization}
The major objective of this thesis is to provide computation based solutions to reduce sensor requirement in advanced model based control techniques for the SPVDG system. Every chapter takes a step forward in this direction and finally culminates into development of nonlinear control techniques with observers to reduce sensors.

The chapter-wise descriptions of the work done in this thesis is given as follows:

\textbf{Chapter \ref{chapter:fastestmppt}} portrays the development of a temperature and irradiance estimation technique that obviates irradiation and temperature sensors in high performance model based MPPT control techniques. 

\textbf{Chapter \ref{chapter:DirectPerturbMPPT}} describes a thrifty and unified control strategy for MPPT and voltage control which is very simple to implement and requires no sensors to be placed on the PV panel. 

\textbf{Chapter \ref{chapter:nonlinear}} delineates a back-stepping based nonlinear control technique which directly uses the large signal model of the SPVDG and provides an extended range of operation along with faster performance.

\textbf{Chapter \ref{chapter:distobs}} incorporates the concept of disturbance observer into the back-stepping control strategy developed in Chapter  \ref{chapter:nonlinear} to reduce the number of sensors used while preserving the benefits of using a nonlinear control scheme.

\textbf{Chapter \ref{sec:conclusion}} concludes the thesis with some observations and indicates the future scope of work to be carried out.

\section{Thesis Contributions}
The major contributions of this thesis can be summarized as follows:
\begin{itemize}
\item For the first time, Newton-Raphson technique has been applied to estimate the PV array curve with only two operating points. 
\item An online technique for estimation of temperature and irradiation considering non-idealities in PV model was developed.
\item An integrated algorithm is developed which exploits the natural electrical interconnection between both the PV and battery subsystems to achieve both MPPT and ancillary service like voltage control in a concomitant manner. This method is very simple with negligible computational burden and carries out two functions in a single control module. It is a low cost method as it uses less sensors.
\item Reduction in transient time while regulating the DC voltage and achieving MPPT of SPVDG system through incorporation of back-stepping based control. The major takeaway from this concept is the manner in which the entire system model was fragmented inorder to successfully apply the nonlinear technique.
\item A disturbance observer was incorporated to estimate different parameters in the SPVDG system to reduce sensors and facilitate the back-stepping based control.
\end{itemize}

%% file: Content/02.tex
       \chapter{Fast Model-based MPPT for PV Arrays with Temperature and Irradiation Estimation }\label{chapter:fastestmppt}
\chaptermark{FAST ESTIMATION BASED MPPT}
\index{Estimation}\index{MPPT!Maximum Power Point Tracking} \index{Sensorless}
 
In this chapter, an algorithm is proposed to achieve maximum power point tracking in a single step under fast varying environmental and load conditions. First, the ambient irradiation and temperature values are estimated using Newton-Raphson (NR) method. Then, the reference PV voltage and current for the corresponding MPP is calculated.  

\section{Introduction} \label{sec:introMPPT} \index{Model-based MPPT}
In the present day scenario, solar photovoltaic energy is considered as a viable alternative to the conventional energy sources such as thermal, gas, nuclear, etc. Therefore, research is being carried out on various issues related to electric power generation for different applications using photovoltaic (PV) energy sources \cite{a1,a2,a3,a4}. PV energy system has several advantages such as pollution-free, abundant availability, less maintenance and near zero carbon emission. However, the non linear current voltage (I-V) characteristic of PV arrays makes it necessary to operate at MPP in order to extract maximum power from it. There are many MPPT algorithms reported in the literature \cite{a6,a7,a8}. Efficient MPPT algorithms should impel PV systems to harvest maximum power available irrespective of the change in atmospheric condition or load. The popular MPPT algorithms are mainly based on different techniques like perturb and observe (P\&O) \cite{a9}\cite{a10}, incremental conductance (INC) \cite{a11}\cite{a12}, hill climbing, fractional open-circuit voltage \cite{a13}, fractional short-circuit current \cite{a14}, ripple correlation control \cite{a15}, fuzzy logic control \cite{a16}, \cite{a17}, particle swarm optimization \cite{a18}, \cite{a19}, artificial neural network \cite{a20}, genetic algorithm \cite{a21}. 
These algorithms differ from one another based on their ease of implementation, tracking speed, number of sensors used, tracking efficiency, cost, etc.  
The P$\&$O MPPT technique is quite straight forward in computation and can easily be implemented using any low-cost microcontroller. However, in steady-state, the output power oscillates around the MPP resulting in inefficient extraction of available power \cite{a10}. These power oscillations can be minimized by reducing the voltage step size, but it takes more time to reach the MPP. Moreover, during rapid change in environmental conditions, there is a possibility that the operating point of PV system may deviate from MPP \cite{a22}.  
\par

In \cite{vsinc}, a variable step INC method (VSINC) was proposed where the voltage step size is adaptively varied based on the rate of change of power with voltage. Even then, as the operating point reaches MPP, the step size reduces to a smaller value which requires more time to converge. However, in this technique also, settling time to reach MPP is high because the voltage step size is reduced to a smaller value as the operating point reached near MPP. Further, a fast converging MPPT (FC-MPPT) method was proposed in \cite{mb2015} which quickly shifts the operating point near to MPP region using geometric techniques and then applies incremental conductance to reach the actual MPP. However, in case of fast varying environmental and load conditions, this method does not perform well because the obtained approximate operating points tend to be quite far from MPP region as there is no information of the system model being used. The MPPT methods discussed so far gradually arrive at MPP by varying the reference voltage/current values based on certain search criteria. They search for MPP in each step without actually possessing the information about complete characteristics of PV system. Alternatively, many model based MPPT techniques have been proposed in literature \cite{annsurvey2015}-\cite{a25}. These techniques capture the complete range of I-V relationship using different models and directly find the MPP using computational means. The PV system is then directly controlled to operate at MPP. Various  model estimation techniques using artificial neural networks (ANN) \cite{annsurvey2015} and neuro-fuzzy models \cite{fuzzyneural1} were proposed for this purpose. While these techniques find the MPPs accurately, they often suffer from heavy computational burden. Moreover, if there are unprecedented changes in environmental and physical conditions, these models need to be retrained to accurately reflect the changes. On the other hand, physical model estimation techniques using curve-fitting methods have also shown much presence in the recent literature. These techniques solve the different electrical models of the solar cell such as single-diode model \cite{singlediode}, double diode-model \cite{doublediode} under controlled conditions and extrapolate them for other working conditions and find the MPP using numerical techniques. Many of these resort to \cite{curvesweep}, analytical five point method \cite{fivepoint} or heuristic techniques like PSO \cite{pso} to arrive at the accurate characteristics of the solar cell. An analytical expression was also developed for the direct determination of MPP references using Lambert W function \cite{a25}. However, in these model estimation based MPPT methods, additional hardware is required for measuring temperature and irradiance which adds to cost of the overall system. 

\par

\textcolor{black}{To eliminate the need of temperature and irradiation sensors, many methods has been developed to estimate the temperature and irradiance and improve detection of the MPP. For instance, \cite{mb2016} proposed a combined model based and heuristic MPPT (CMH-MPPT) which works similar to FC-MPPT method by forcing the operating point to a near MPP zone. However, its performance is superior to that of FC-MPPT during rapid change in environmental conditions due to the additional model based temperature estimation feature included. The limitation of this method is that it cannot reach MPP in a single step. On the contrary, \cite{directdcalc2017} proposed a single-step formula for finding MPP using simplified polynomial analytical model which estimates its parameters in real time without using additional environmental sensors. However, its single-step formula necessitates placement of an additional voltage sensor at the output. It is also observed that in most of these works, the non-idealities of PV models have been neglected for simplicity. To the best of the authors' knowledge, there is no work in the literature to find MPP for fast varying environmental and load conditions in a single step using only voltage and current sensors while considering PV array non-idealities.}
\par

\index{Single Step MPPT}

In this chapter, an algorithm is proposed to achieve maximum power point tracking in a single step under fast varying environmental and load conditions. First, the ambient irradiation and temperature values are estimated using Newton-Raphson (NR) method. Then, the reference PV voltage and current for the corresponding MPP is calculated.

The following is the organization of this chapter. Section \ref{sec:tempestmodel} elucidates the mathematical modeling of a PV array and PV array characteristics. Section  \ref{sec:tempestimation} discusses the temperature and irradiation technique proposed in this chapter using only two operating points on the curve. Section \ref{sec:tempestdesstate} shows how to obtain the desired values of all the states present in the PV array system. Section \ref{sec:tempestresult} shows the estimation results that are obtained in different temperature and irradiation conditions and also in different loading conditions while Section \ref{sec:tempestsummary} summarizes the entire technique and its application. 

\section{Mathematical Modeling of the PV Array}\label{sec:tempestmodel}
\index{PV Model}
\textcolor{black}{The basic PV system configuration and its mathematical description is presented in the following subsections:}
\subsection{PV characteristic}
\begin{figure}
\centering
\includegraphics[width=0.8\linewidth]{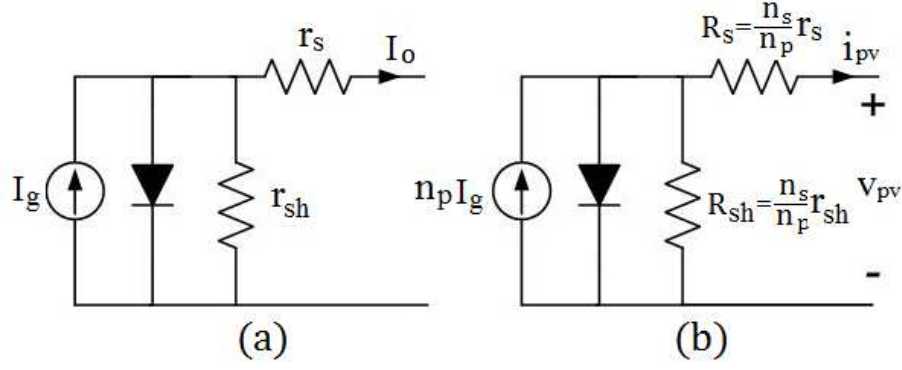}
\caption{Equivalent circuit of (a) Single PV cell (b) Complete PV array}
\label{cell1}
\end{figure}

The electrical characteristics of a PV cell are usually described by the single-diode model with acceptable accuracy \cite{a4.5}.  The equivalent circuit of a single PV cell using single-diode model is shown in Fig.\ref{cell1}(a). In practice, PV cells are combined in series and parallel to form a large PV array. The equivalent circuit of a PV array consisting of $n_{s}$ series and $n_{p}$ parallel PV cells is shown in Fig.\ref{cell1}(b). The current-voltage relationship of the PV array is given by \cite{a32iveqn}
\begin{align}\label{ipv}
\scalebox{.9}{$
i_{pv}=n_{p}I_{g}-n_{p}I_{s}\left(e^{\dfrac{q(v_{pv}+i_{pv}R_{s})}{n_{s}pKT}}-1\right)-\dfrac{v_{pv}+i_{pv}R_{s}}{R_{sh}}$}
\end{align}

where \\
$~~~~~~~~~~~I_g$: Photogenerated current of a single PV cell \\
$~~~~~~~~~~~ i_{pv}$: Output current of the PV array \\
$~~~~~~~~~~~v_{pv}$: Output voltage of the PV array \\
$~~~~~~~~~~~I_{s}$: Reverse saturation current of diode\\
$~~~~~~~~~~~r_s$: Series resistance of a single cell \\
$~~~~~~~~~~~r_{sh}$: Parallel resistance of a single cell \\
$~~~~~~~~~~~n_{s}$: Numbers of series cells in PV array \\
$~~~~~~~~~~~n_{p}$: Numbers of parallel cells in PV array \\
$~~~~~~~~~~~R_s$ Series resistance of PV array \\
$~~~~~~~~~~~R_{sh}$: Parallel resistance of PV array \\
$~~~~~~~~~~~p$: Diode ideality constant \\
$~~~~~~~~~~~T$: Actual cell temperature \\
$~~~~~~~~~~~T_{r}$: Reference cell temperature (298 $K$) \\
$~~~~~~~~~~~\lambda$: Actual irradiance \\
$~~~~~~~~~~~\lambda_{r}$: Reference irradiance (1000 $W/m^2$) \\
$~~~~~~~~~~~k_{i}$: Short-circuit current temperature coefficient. \\
$~~~~~~~~~~~I_{sc}$: Short-circuit current at STC  (298 $K$ $\&$ 1000 $W/m^2$ \\
$~~~~~~~~~~~q$: Electron charge  ($1.6\times 10^{-19} ~C$) \\
$~~~~~~~~~~~K$: Boltzmann constant  ($1.38\times 10^{-23} J/K$\\
$~~~~~~~~~~~E_{gp}$: Energy bandgap  ($1.1 ~eV$) 

The photo-generated current $I_{g}$ is a function of temperature and irradiance and is given as 
\vspace{-.1cm}
\begin{align}\label{ig}
I_{g} = (I_{sc} + k_{I}(T-Tr))\lambda/\lambda_{r}
\end{align}

\hspace{-.4cm}The reverse saturation current $I_{s}$ is formulated as 
\vspace{-.4cm}

\begin{align}\label{is}
I_{s} = I_{r}  \left( \frac{T}{T_r}\right)^{3} e^{qE_{gp}\left(\frac{1}{T_r}-\frac{1}{T}\right)/pK}
\end{align}

\begin{figure}
\centering
\includegraphics[width=0.8\linewidth]{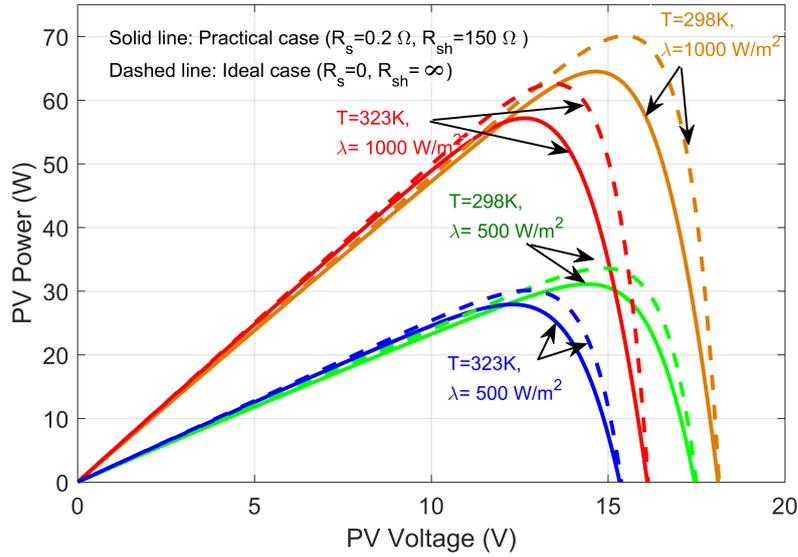}
\caption{Effect of series and shunt resistance of P-V curve for different temperature and irradiance.}
\label{pvchar}
\end{figure}

As discussed earlier, the variations in irradiance and temperature affect the power output of the solar PV array. Similarly, series and shunt resistances also affect the output power. These effects have been portrayed in the P-V characteristics as shown in Fig.\ref{pvchar}. It may be observed that although 
$R_{s}$ and $R_{sh}$ have least effect in open circuit and short circuit region, they greatly influence the P-V characteristic in MPP region. Therefore, it is important to take them into consideration while estimating actual $T$ and $\lambda$ using I-V curve.\textcolor{black}{Although $R_{s}$ and $R_{sh}$ vary with $T ~\& ~\lambda$ \cite{rs1}, \cite{rs2}, their variation do not cause significant effect on I-V and P-V characteristic as far as $T ~\& ~\lambda$ are in operating region. Therefore, we have assumed $R_{s}$ and $R_{sh}$ to be constant in our method.}

\subsection{Power Converter}
\begin{figure}
\centering
\includegraphics[width=0.8\linewidth]{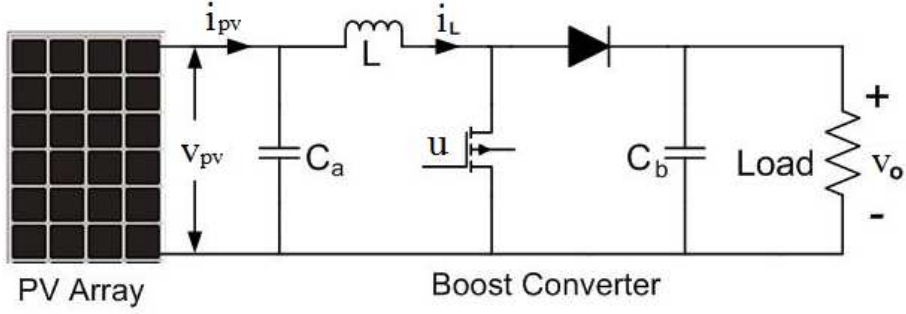}
\caption{PV Array in conjunction with dc-dc boost converter.}
\label{converter}
\end{figure}

Fig. \ref{converter} shows a PV array connected to a dc-dc boost converter. Power output from PV module is controlled by varying the duty ratio of boost converter so as to operate at MPP. The PV system along with boost converter is a third order non-linear system. Taking inductor current $i_{L}$, PV voltage $v_{pv}$ and load voltage $v_{o}$ as system states, the time averaged state space model of boost converter \cite{haroun} is described as follows,

\vspace{-.25cm}
\begin{align}\label{ildot}
\dot{i}_{L}= \dfrac{1}{L}\left(v_{pv}-ri_{L}-V_{D}-v_{o}\right)+\dfrac{1}{L}(V_{D}+v_{o})u
\end{align}
\vspace{-.4cm}
\begin{align}\label{vpvdot}
\dot{v}_{pv}=\dfrac{1}{C_{a}}\left(i_{pv}-i_{L} \right)
\end{align}
\vspace{-.4cm}
\begin{align}\label{vodot}
\dot{v}_{o}=\dfrac{1}{C_{b}}\left( i_{L}- \dfrac{v_{o}}{R_{ld}}\right)-\dfrac{1}{C_{b}}i_{L}u 
\end{align}

\textcolor{black}{Here $C_{a}:$ PV capacitor, $C_{b}:$ output capacitor, $V_{D}:$ diode voltage drop, $r:$ resistance of inductor, $R_{ld}:$ load resistance. $u$ is a switching signal which is 1 when switch is on and 0 when switch is off.}

\section{Temperature and Irradiance Detection }\label{sec:tempestimation}
\index{Temperature Detection} \index{Irradiance Detection}

The I-V characteristics of a particular PV array depends on a number of parameters as described in equations (\ref{ipv}) to (\ref{is}). Temperature and irradiance level can be considered as external parameters which change with environmental condition. Other parameters like ideality factor and modeled resistances of PV array are considered as internal parameters which are known and constant. Therefore, variation in I-V characteristic corresponds to the variation in PV array external parameters.
These parameters can be calculated by plugging in different sets of sensed PV array voltage and current values in system equation. Since the I-V characteristic follows a non linear relation, NR method is used to find the
system parameter as it converges faster than other techniques.    
\par

The I-V relation can be written as function of internal and external parameters as
\begin{align*}
i_{pv}=f({\textbf{x}},T,\lambda , v_{pv})
\end{align*}

where ${\textbf{x}}$ is set of internal parameters of PV array which are considered constant. \par 

\index{Newton-Raphson}
Following the prescripts of NR Method, it is assumed that $(v_{1},i_{1})$ and $(v_{2},i_{2})$ are two sets of measured value of PV array voltage and current at time t and $t+\delta t$ while operating at temperature $T$ and irradiance $\lambda$. $\delta t$ is the time interval at which temperature and irradiance are updated. Hence, $i_{pv_1}$ and $i_{pv_2}$ at $v_{pv_1}$ and $v_{pv_2}$ can be written as
\begin{align*}
i_{pv_1}=f({\textbf{x}},T,\lambda , v_{pv_1})     \hspace{1cm} i_{pv_2}=f({\textbf{x}},T,\lambda , v_{pv_2})
\end{align*}

It is also assumed that $i_{pv_{1k}}$ and $i_{pv_{2k}}$ are the calculated currents at voltages $v_{pv_11}$ and $v_{pv_2}$ respectively in the $k^{th}$ iteration corresponding to estimated temperature $T_{k}$ and irradiance $\lambda_{k}$.
Therefore,
\begin{align*}
i_{pv_{1k}}=f({\textbf{x}},T_{k},\lambda_{k} , v_{pv_1})\hspace{1cm}   i_{pv_{2k}}=f({\textbf{x}},T_{k},\lambda_{k} , v_{pv_2})
\end{align*}

Having known the internal parameters and sets of measured values of PV voltage and current, temperature $T_{k+1}$ and irradiance $\lambda_{k+1}$ in the next iteration is calculated as follows \par 

\begin{align}
\begin{bmatrix}
    T_{k+1}       \\
    \lambda_{k+1}       
\end{bmatrix}
=
\begin{bmatrix}
    T_{k}       \\
    \lambda_{k} 
\end{bmatrix} 
+ \hspace{.1cm}
[J]^{-1}
\hspace{.1mm}
\begin{bmatrix}
    i_{pv_1}-i_{pv_{1k}}       \\
     i_{pv_2}-i_{pv_{2k}} 
\end{bmatrix}
\end{align}
where
\begin{align}
[J]=\begin{bmatrix}
\dfrac{\partial i_{pv}}{\partial T}_{(v_{pv_1},T_{k},\lambda _{k})} & \dfrac{\partial i_{pv}}{\partial \lambda}_{(v_{pv_1},T_{k},\lambda _{k})}\\
     \dfrac{\partial i_{pv}}{\partial T}_{(v_{pv_2},T_{k},\lambda _{k})} & \dfrac{\partial i_{pv}}{\partial \lambda}_{(v_{pv_2},T_{k},\lambda _{k})} 
\end{bmatrix}
\end{align}
 $\dfrac{\partial i_{pv}}{\partial T}$ is found out by differentiating (\ref{ipv}) and then upon simplification is written as \par

\begin{align}
 \dfrac{\partial i_{pv}}{\partial T} = n_{p}\dfrac{\partial I_{g}}{\partial T}-n_{p}\dfrac{\partial I_{s}}{\partial T}\left(e^{\dfrac{q(v_{pv}+i_{pv}R_{s})}{pKn_{s}T}}-1\right)- 
 \hspace{4mm} \\        \hspace*{1mm}n_{p}I_{s}\dfrac{\partial}{\partial T}\left(e^{\dfrac{q(v_{pv}+i_{pv}R_{s})}{pKn_{s}T}}-1\right)-
\dfrac{\partial}{\partial T}\dfrac{(v_{pv}+i_{pv}R_{s})}{R_{sh}} \nonumber
\end{align}
Taking the terms of $\dfrac{\partial i_{pv}}{\partial T}$ together, we get,
\vspace{-.2cm}

\vspace{-0.4cm}
\begin{figure*}[!h]
\begin{equation}
\resizebox{0.8\textwidth}{!}{$
\dfrac{\partial i_{pv}}{\partial T} = \dfrac{n_{p}k_{I}\dfrac{\lambda}{\lambda_{r}}-n_{p}\dfrac{\partial I_{s}}{\partial T}\left(e^{\dfrac{q(v_{pv}+i_{pv}R_{s})}{pKn_{s}T}}-1\right)+ \dfrac{n_{p}I_{s}q(v_{pv}+i_{pv}R_{s})}{pKn_{s}T^2}\left(e^{\dfrac{q(v_{pv}+i_{pv}R_{s})}{pKn_{s}T}} \right)} 
{1+\dfrac{n_{p}I_{s}qR_{s}}{pKn_{s}T}\left(e^{\dfrac{q(v_{pv}+i_{pv}R_{s})}{pKn_{s}T}} \right)+\dfrac{R_{s}}{R_{sh}}}$}  
\end{equation} 
\noindent\rule{18cm}{0.4pt}
\end{figure*}
\begin{align}
\dfrac{\partial I_{s}}{\partial T}=\dfrac{I_{r}}{T_{r}^3}\left(3T^2+\dfrac{qE_{gp}T}{pK}\right)e^{\dfrac{qE_{gp}}{pK}\left(\dfrac{1}{T}-\dfrac{1}{T_{r}}\right)}
\end{align}
\\
Similarly $\dfrac{\partial i_{pv}}{\partial \lambda}$ can be written as

\begin{align}
\dfrac{\partial i_{pv}}{\partial \lambda}=\dfrac{n_{p}\left(I_{sc}+k_{I}(T-T_{r}\right))}{\lambda_{r}+\dfrac{\lambda_{r}n_{p}I_{s}qR_{s}}{pKTn_{s}}\left(e^{\dfrac{q(v_{pv}+i_{pv}R_{s})}{pKn_{s}T}} \right)}
\end{align}

NR method converges very quickly and it typically takes 4 to 5 iterations to find out the solution of $T$ and $\lambda$. The interval $\delta t$ is also an important parameter which should be carefully chosen. While higher value of $\delta t$ makes the detection slower in case of change in environmental conditions, smaller value repetitively calculates the same T and $\lambda$ with high frequency in case of no environmental change. 
\textcolor{black}{To overcome this situation, in this work, the new value of $T$ and $\lambda$ is estimated only when change in PV voltage/current is greater than a minimum value.}

\section{Calculation of desired states}\label{sec:tempestdesstate}
Desired states of the system are calculated using state space equations once the PV characteristic is known. PV voltage $v_{pv_{r}}$ and current $i_{pv_{r}}$ corresponding to MPP are numerically found out using INC method as done in \cite{smc}. Calculation of desired value of inductor current and converter output voltage is performed using steady state analysis. 

\begin{table}
\caption{Desired values of PV voltage, Current and Power for different environmental conditions}
\vspace{-.5cm}
\label{desired}
\begin{center}
\begin{tabular}{|cccccc|}
\hline
Case &Temperature &Irradiance& $v_{pv_{r}}$ &$i_{pv_{r}}$&$P_{pv_{r}}$\\
\hline
I&298K&500$W/m^{2}$&14.4V&2.15A&31.0W\\

II&298K&1000$W/m^{2}$&14.6V&4.40A&64.2W\\

III&323K&1000$W/m^{2}$&12.7V&4.52A&57.5W\\

IV&323K&500$W/m^{2}$&12.3V&2.26A&28.5W\\
\hline
\end{tabular}
\vspace{-.5cm}
\end{center}
\end{table}

\par 
Let $i_{L_{r}}$, $v_{pv_{r}}$, $v_{o_{r}}$ be the desired steady-state values of inductor current, PV voltage and converter output voltage respectively and $e_{1}=i_{L}-i_{L_{r}}$, $e_{2}=v_{pv}-v_{pv_{r}}$, $e_{3}=v_{o}-v_{o_{r}}$ be be their corresponding errors. In steady state, when the desired values are reached, all the errors and their derivatives become 0.
\begin{align} \label{errors}
\setcounter{equation}{12}
e_{1}=e_{2}=e_{3}=0,\dot{e}_{1}=\dot{e}_{2}=\dot{e}_{3}=0
\end{align}
Therefore, in steady state, \eqref{ildot} to \eqref{vodot} become
\begin{align}\label{ildot1}
\dfrac{1}{L}(v_{pv_{r}}-ri_{L_{r}}-V_{D}+u_rV_{D}-v_{o_{r}}+u_rv_{o_{r}})=0
\end{align}
\vspace{-.5cm}
\begin{align}\label{vpvdot1}
\dfrac{1}{C_{a}}(i_{pv_{r}}-i_{L_{r}})=0
\end{align}
\vspace{-.5cm}
\begin{align}\label{vodot1}
\dfrac{1}{C_{b}}\left(i_{L_{r}}(1-u_{r})-\dfrac{v_{o_{r}}}{R_{ld}}\right)=0
\end{align}


Upon simplifying, \eqref{ildot1}, \eqref{vpvdot1} and \eqref{vodot1} we get,
\begin{align}\label{ildot2}
i_{L_{r}}=i_{pv_{r}}
\end{align}
\vspace{-.4cm}
\begin{align}\label{vpvdot2}
v_{o_{r}}=\dfrac{-V_{D}+\sqrt{V_{D}^2+4i_{pv_{r}}R_{ld}(v_{pv_{r}}-ri_{pv_{r}})}}{2}
\end{align}
\vspace{-.5cm}
\begin{align}\label{vodot2}
u_{r}=1-\dfrac{v_{pv_{r}}-i_{pv_{r}}r}{V_{D}+v_{o_{r}}}
\end{align}

where $u_r$ is the desired duy cycle.
\par
Equations \eqref{ildot2} to \eqref{vodot2} show that the desired steady state inductor current is equal to the reference PV current corresponding to MPP. Although desired PV current for MPP and hence desired average inductor is independent of load, the final output voltage is a function of load resistance which makes the desired duty cycle dependent on load. Table \ref{desired} shows desired values of MPP PV voltage, current and power for different temperature and irradiance levels. These desired states can be achieved by a range of controllers which include different varieties of linear and non-linear techniques.
\section{Simulation Results and Discussion}\label{sec:tempestresult}
\begin{figure}
\centering
\includegraphics[width=0.8\linewidth]{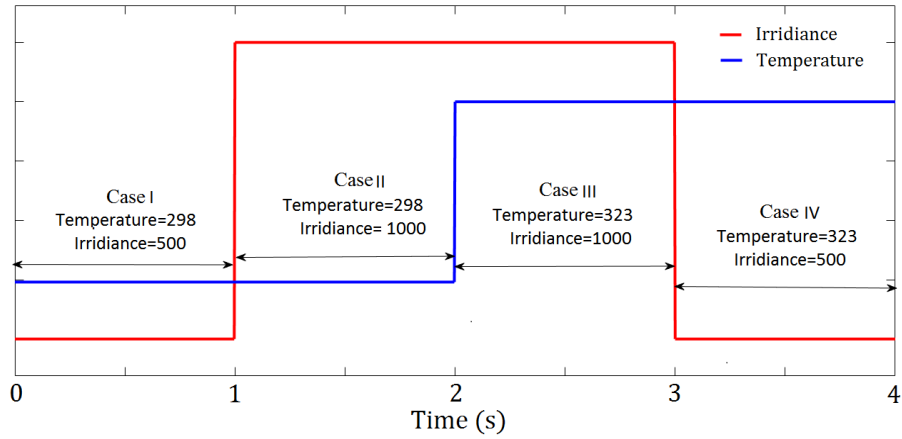}
\caption{Different cases of variation in temperature and irradiance}
\label{reg}
\end{figure}

\begin{figure}
\centering
\includegraphics[width=0.8\linewidth]{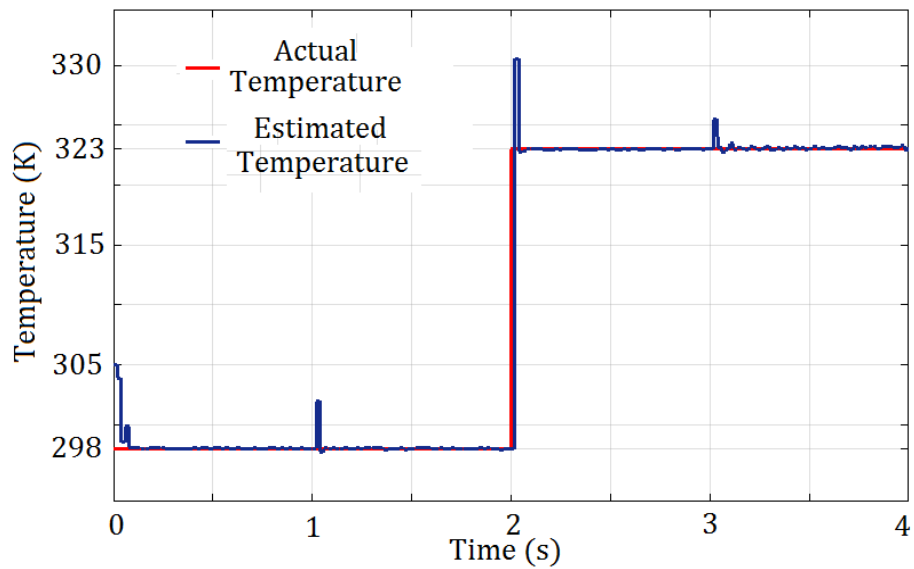}
\caption{Comparison of estimated temperature with actual value}
\label{tempest}
\end{figure}

\begin{figure}
\centering
\includegraphics[width=0.8\linewidth]{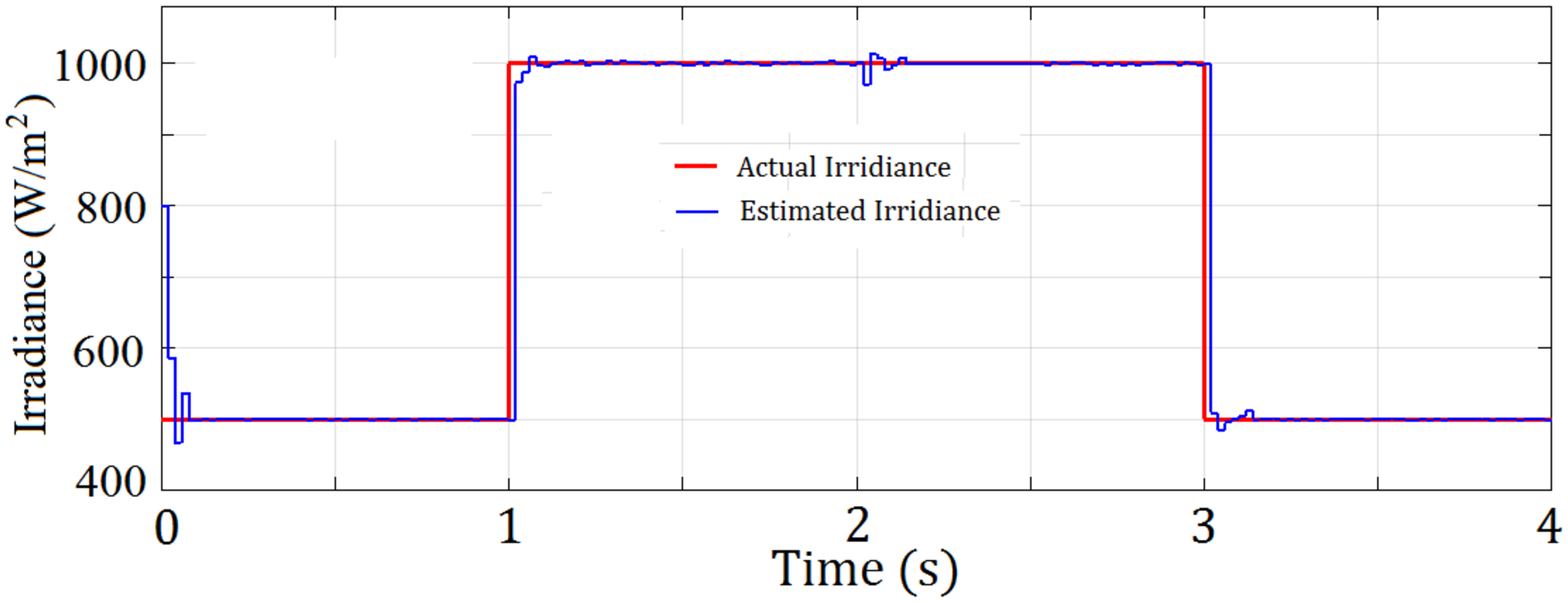}
\caption{Comparison of estimated irradiance with actual value}
\label{radest}
\end{figure}

The performance of the proposed algorithm is simulated using MATLAB 2016a software on Intel core i3, 2.4GHz processor, 4GB RAM and Windows 7 operating system. PV array block has been used to simulate solar array. The operating temperature and irradiance can be given as external input and varied as desired. The parameters of the PV array block is shown in Table \ref{parameters}.
The pattern of variation in temperature and irradiance is shown in Fig. \ref{reg}. The reference PV voltage and current values are given in Table \ref{desired} for these four cases. The estimated value of temperature and irradiance are obtained in Fig. \ref{tempest} and Fig. \ref{radest} respectively using the procedure discussed in Section \ref{sec:tempestdesstate}. It can be observed that the variation of estimated values are within $0.5\%$ of their actual values. These values are estimated at the interval of 0.05$s$. It is also observed that the two parameters reach their steady state value within 0.25$s$. Although temperature is changed at $t=2s$, there are some transients in estimated temperature at $t=1s$ and $t=3s$. Similarly transients are observed in irradiance estimation at $t=2s$ when it is changed at $t=1s$ and $t=3s$. These transients are due to the fact that the calculation of one parameter is not independent from another. The P-V curve is changed if any of these two parameters is changed. These parameters cannot be correctly calculated if the two sets of voltage and currents are from different P-V curves. Once the P-V characteristic settles on a single curve, the parameter estimation is correctly done. 

 \begin{table}
 \caption{parameters used for simulation and hardware implementation}
 \vspace{-.5cm}
 \label{parameters}
 \begin{center}
 \begin{tabular}{ |p{1.5cm}p{1.5cm}|p{1.5cm}p{1.5cm}|  }
 \hline
 \multicolumn{2}{|c|}{PV array parameter} & \multicolumn{2}{|c|}{Circuit parameter} \\
 \hline
 
 \hspace{.5cm}$T_{r}$ &   298K                              &   \hspace{.3cm}  $C_{a}$ & $200 \mu F$ \\
 \hspace{.5cm}$\lambda_{r}$ & 1000$W/m^{2}$                 & \hspace{.3cm} $C_{b}$ & $200 \mu F$ \\
 \hspace{.5cm}$p$      &   1                                  & \hspace{.3cm} $L$   & 5mH   \\
 \hspace{.5cm}$I_{r}$ & $1.37\hspace{-0.2mm}$x$10^{-8}$A     &  \hspace{.3cm} $r$   & $0.2\Omega$\\
 \hspace{.5cm}$I_{sc}$ & $4.8$A                             & \hspace{.3cm} $V_{d}$&  0.6V\\\cline{3-4}
\hspace{.5cm}$R_{s}$&   $0.2 \Omega$\   &\multicolumn{2}{|c|}{Controller parameter}\\\cline{3-4}
  \hspace{.5cm}$R_{sh}$& $150 \Omega$                         & \hspace{.3cm} $f_{s}$    & 20KHz\\
 \hspace{.5cm}$n_{s}$& 36                                 &  \hspace{.3cm} $K_{c}$   & 1.5 \\
 \hspace{.5cm}$n_{p}$& 1                             &  \hspace{.3cm} $t_{o}$   & 0.025s\\
 &                                     &  \hspace{.3cm}  $\partial t$  & 0.05s    \\
 \hline
\end{tabular}
\end{center}
\end{table} 

\begin{figure}
\centering
\includegraphics[width=0.8\linewidth]{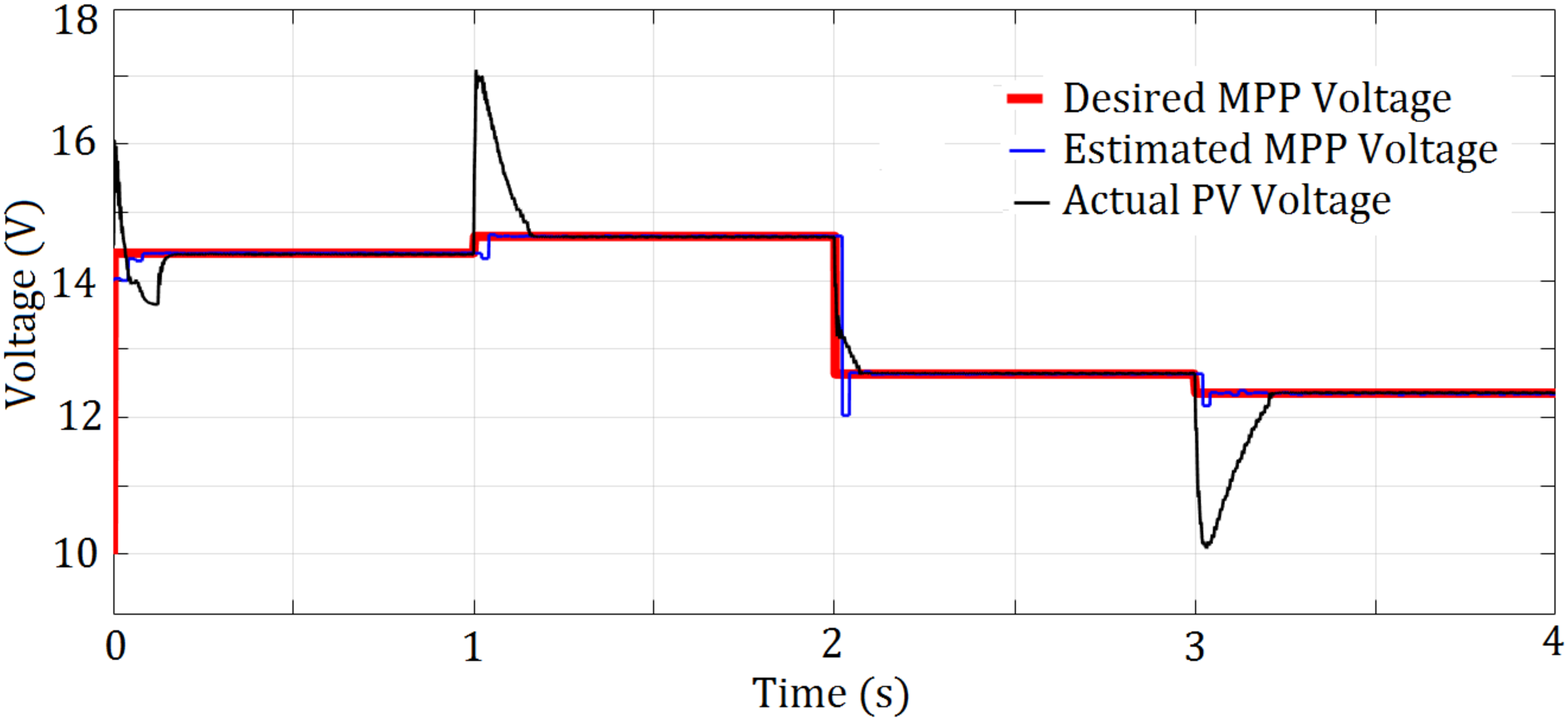}
\caption{Desired voltage, estimated voltage and actual current for MPP under different environmental conditions.}
\label{voltage}
\end{figure}

\begin{figure}
\centering
\includegraphics[width=0.8\linewidth]{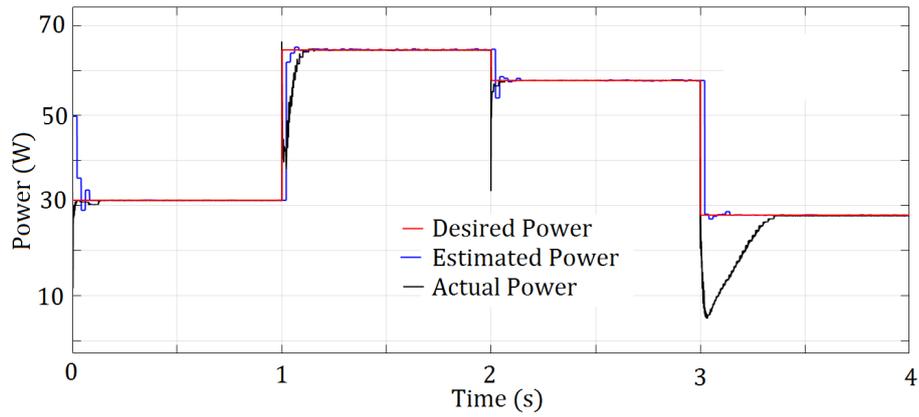}
\caption{Desired Power, estimated power and actual power for MPP under different environmental conditions.}
\label{power}
\end{figure}

\begin{figure}
\centering
\includegraphics[width=0.8\linewidth]{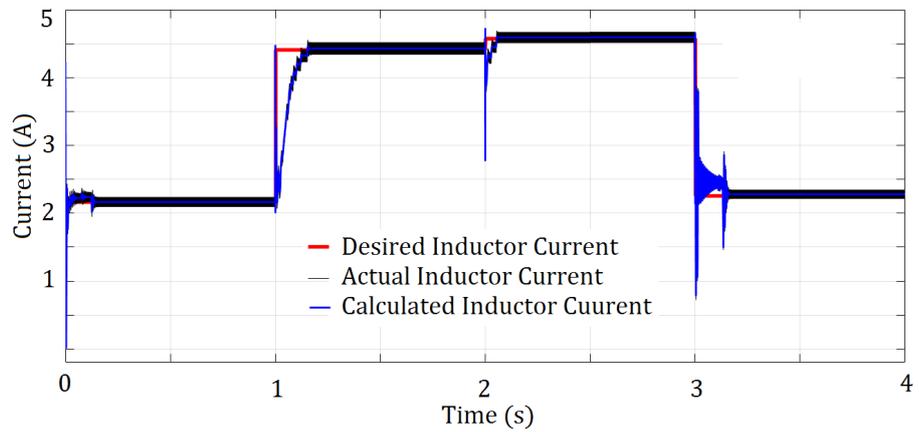}
\caption{Desired, calculated and actual inductor current at MPP for different environmental conditions}
\label{current}
\end{figure}

\begin{figure}
\centering
\includegraphics[width=0.8\linewidth]{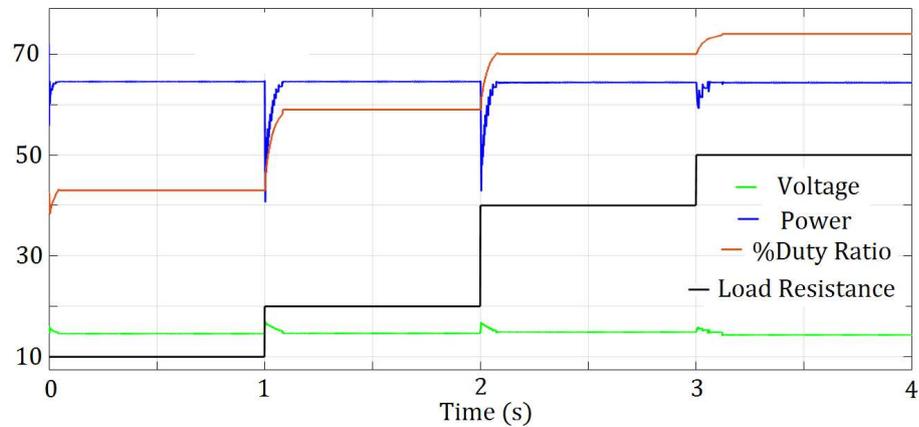}
\caption{PV voltage, power and duty ratio for different load conditions}
\label{resistance}
\end{figure}

Figures \ref{voltage} and \ref{power} show the dynamics of desired, estimated and actual values of PV voltage and PV power respectively. Temperature and irradiance of PV array block are changed as per Fig. \ref{reg}. Time taken to reach steady state values is 0.05$s$. The desired, calculated and actual inductor current is shown in Fig. \ref{current}. It is to be noted that calculated inductor current tracks the actual current. The steady state ripple in calculated inductor current is much lower compared to its actual value because the calculated inductor current is function of PV voltage and PV current which are constant in steady state. 
However, these transients are present in calculated inductor current whenever there is variation in operating conditions. The transient time depends on the input capacitor value. Fig.\ref{resistance} shows the variation in PV voltage, power and duty ratio when load is changed.

\section{Summary} \label{sec:tempestsummary}
This chapter presents a fast MPPT technique for photovoltaic array by estimating environmental conditions and load. For temperature and irradiance estimation, Newton-Raphson method has been applied to I-V characteristic of the PV array. Load was estimated by calculating inductor current and load voltage using boost converter dynamics. Simulation results show that the proposed method is able to operate the PV array at MPP under varying environmental and load conditions. The proposed method is able to find the value of duty cycle in single step for MPPT which otherwise is gradually found. This enables the proposed method to be free from multiple transients during MPPT and steady state oscillation. Application of proposed method to the case of partial shading will be carried out in the future.

%% file: Content/03.tex
\chapter{A Unified Sensor-free Control Strategy for MPPT and Voltage Control in Standalone PVDG Systems}\label{chapter:DirectPerturbMPPT}\index{Direct Perturbation MPPT} 
\chaptermark{UNIFIED MPPT AND VOLTAGE CONTROL}
In this chapter, a description is given on developing a unified control approach for mppt and voltage control without placing any sensors near the PV array. This has been achieved keeping in mind the power balance conditions of the Standalone PVDG system.

\section{Introduction }\index{IntroMGStandalone}

In SPVDGs, DC bus voltage control and MPPT are generally carried out separately. An extensive treatment of various MPPT techniques has already been done in the previous chapter and the voltage control carried out by many linear and nonlinear approaches with different higher level control structures like autonomous control, coordination control and supervisory control \cite{uniref1}.  All of these methods require different sets of voltage and current sensors for both the control tasks of the SPVDG namely MPPT and DC voltage control. 

Inorder to scale these technologies to the underprivileged  parts of the world where SPVDGs are actually required, any technique which would reduce the cost of the system is very much welcome. By developing intelligent methods to reduce sensors, it is possible to reduce the overall cost of the system. The concept of sensor reduction in SPVDG systems is hardly explored in literature although it had been very much considered in drive based systems. For instance, \cite{uniref2} uses observers to remove the current sensor on the PV array side necessary for tracking MPPT. \cite{uniref3} proposes using only averaged voltages of a photovoltaic array and the average duty ratios to eliminate the current sensor in the MPPT process.\cite{uniref4} discusses how to use only one sensor in ESS control for DC voltage stabilization in SPVDG system. However, none of them provide any solution to carry out both MPPT and voltage control with reduced sensors.

With these concerns, this chapter proposes a direct perturbation based MPPT algorithm for a standalone DC microgrid which requires no sensor to be placed with the PV array side. A set of logic conditions are derived with respect to any change in grid voltage. The conditions are processed within an algorithm to decide the next perturbation for reaching MPP. More importantly, this single algorithm not only serves for MPPT but also accomplishes DC bus voltage control which is an ancillary service in a conjoint manner.
Automatic battery charging and discharging is carried out through seamless transition between buck and boost modes of operation based on grid voltage operating condition. This is facilitated by a bidirectional converter (BDC) used for integrating BESS to the DC grid which enables two way power flow from battery to grid and vice-versa. It uses the duty ratios for BDC by generating them previously MPP section. 

In essence, a unique and indirect method to achieve MPPT and DC grid voltage control without any sensors connected to PV array for MPPT has been presented in this chapter. An integrated algorithm is developed which exploits the natural electrical interconnection between both the PV and battery subsystems to achieve both MPPT and ancillary service like voltage control in a concomitant manner. This method is very simple with negligible computational burden and carries out two functions in a single control module. It is a low cost method as it uses less sensors.

The following is the organization of this chapter. Section \ref{sec:unimodel} describes the SPVDG system under consideration and its working. Section \ref{sec:unicontrol} gives a comprehensive understanding of the direct perturbation based unified MPPT and voltage control technique. Section \ref{sec:uniresults} shows the results that are obtained for the MATLAB/Simulink simulations while Section \ref{sec:unisummary} concludes the chapter. 

\section{Configuration Of the SPVDG} \label{sec:unimodel} 
\begin{figure}[!ht]
\centering
\includegraphics[width=0.8\linewidth]{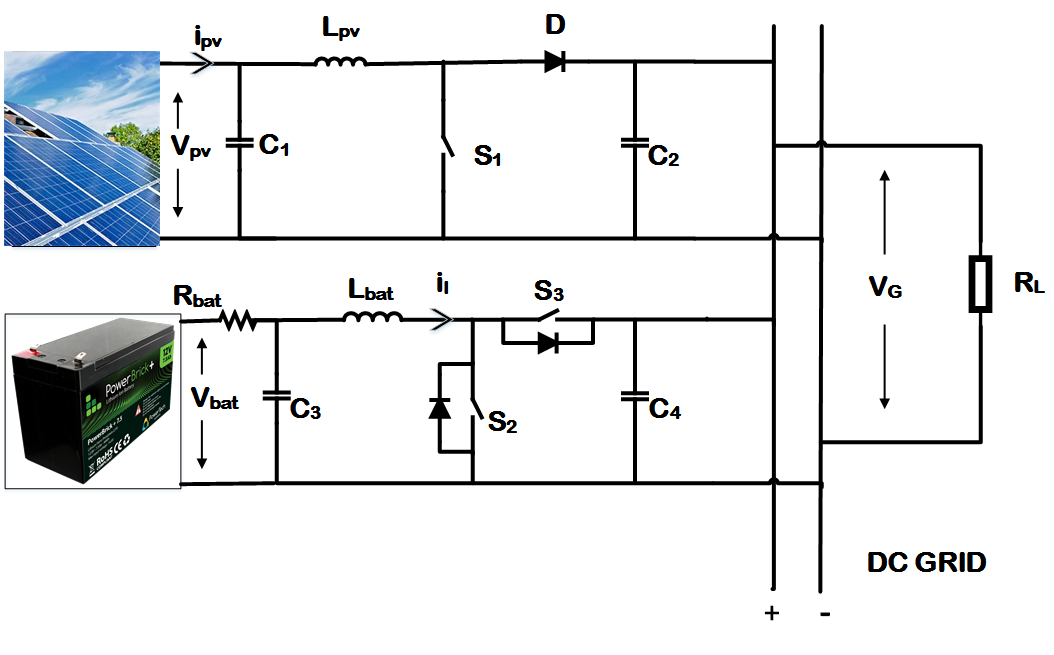}
\caption{Overall schematic of the standalone DC microgrid.}
\label{fig:ms}
\end{figure}

\index{Unified Algorithm} 
A solar photovoltaic based low voltage SPVDG unit is proposed for this work. It operates off-grid with local generation from solar PV and feeds local DC loads. It basically contains both DC/DC boost and BDC with a battery storage device and resistive load. A complete circuit model of the proposed system is given in Fig.\ref{fig:ms}. 

 The PV source is simulated by taking SLX060 USC solar module from simulink library. It generates photovotaic current ($i_{pv}$) and voltage ($V_{pv}$) to meet up with the maximum power rating. A PV cell is the very basic element of an array comprises of a p-n junction diode which converts the solar irradiation into electric energy. The solar module contains 36 number of series connected cells to generate required amount of  PV output voltage. 

\begin{figure}[!ht]
\centering
\includegraphics[width=0.8\linewidth]{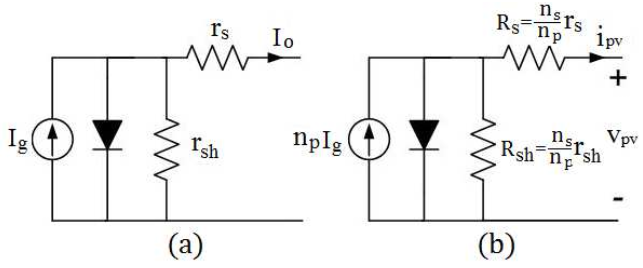}
\caption{Equivalent circuit of (a) single PV cell (b) PV array.}
\label{fig:solar}
\end{figure}

The equivalent circuit of PV cell and an array can be seen in Fig. \ref{fig:solar} where $n_s$ and $n_p$ are number of cells in series and parallel respectively.\\

A DC/DC boost converter boosts up the PV voltage  to a suitable DC reference value at the DC bus through a MPPT algorithm. It is properly modeled with accurate parameters for a unidirectional power flow from PV source to grid \cite{r23}. The switching signal for proper duty cycle is computed by proposed direct perturbation based MPPT algorithm. The output pulse is given to gate of the switch to produce required $V_G$ to be appeared at DC load bus.

A DC/DC bi-directional converter is used to integrate a BESS battery to the DC grid. It enables bi-directional power flow from grid to battery and vice-versa to establish automatic battery charging and discharging by eventually controlling the DC grid voltage \cite{r24}. Dynamic values of duty cycles for both buck and boost operation are calculated by the proposed control algorithm and fed to the respective gate port of the switches. Both the converters use MOSFET switches($S_1$, $S_2$, $S_3$) for their characteristics of being used in low voltage applications (10-100 volts, 10-100 KHz switching frequency).

The BESS consists of a Li-ion battery and is based on the  dynamic run time characteristics such as non-linear open circuit voltage, current dependency of storage duration and effective response to transients. A dynamic resistive load is connected at output considered as the point of DC grid because of the absence of synchronization and frequency related issues in case of the DC system on study. Parameters for all the sub systems and elements are given in Table \ref{tab1}.

\begin{table}[htbp]
\caption{Parameter Specification}
\begin{center}
\begin{tabular}{|c|c|}
\hline
\textbf{Subsystem} & \textbf{\textit{Parameter Specification}} \\
\hline
PV Array at STC & Model: SLX 230 USC module \\ 
& $V_{OC}=16V$, $I_{SC}=4.8A$ \\
& $V_{MPP}=14.6V$, $I_{MPP}=4.4A$ \\  
\hline
Battery & Model: Lithium Ion: $15 V$\\
\hline
DC/DC Converter & $C_1=100\mu F$, $C_2=500\mu F $, $L_{pv}=0.35 mH$ \\
& $f_{sw}=10kHz$ for $S_1$ \\
\hline
DC/DC Bidirectional & $C_3=100\mu F$, $R_{bat}=0.3\Omega $, $L_{bat}=0.3 mH$ \\
Converter& $f_{sw}=10kHz$ for $S_2$ and $S_3$  \\
\hline
PI &  $K_{p_v}$ = 0.2, $K_{i_v}$ = 0.1 \\
Gains & $K_{p_i}$ = 0.2, $K_{i_i}$ = 0.01\\
\hline
\end{tabular}
\label{tab1}
\end{center}
\end{table}
\section{Unified Sensorless Control Strategy}\label{sec:unicontrol}\index{Sensorless}
The proposed control strategy focuses on developing an efficient algorithm which governs both of the following tasks:
\begin{enumerate}
    \item Direct perturbation based sensor-free MPP tracking and
    \item Automatic battery charging/discharging with seamless mode transition for DC grid voltage control.
\end{enumerate}

\begin{figure}[!ht]
\centering
\includegraphics[width=\linewidth]{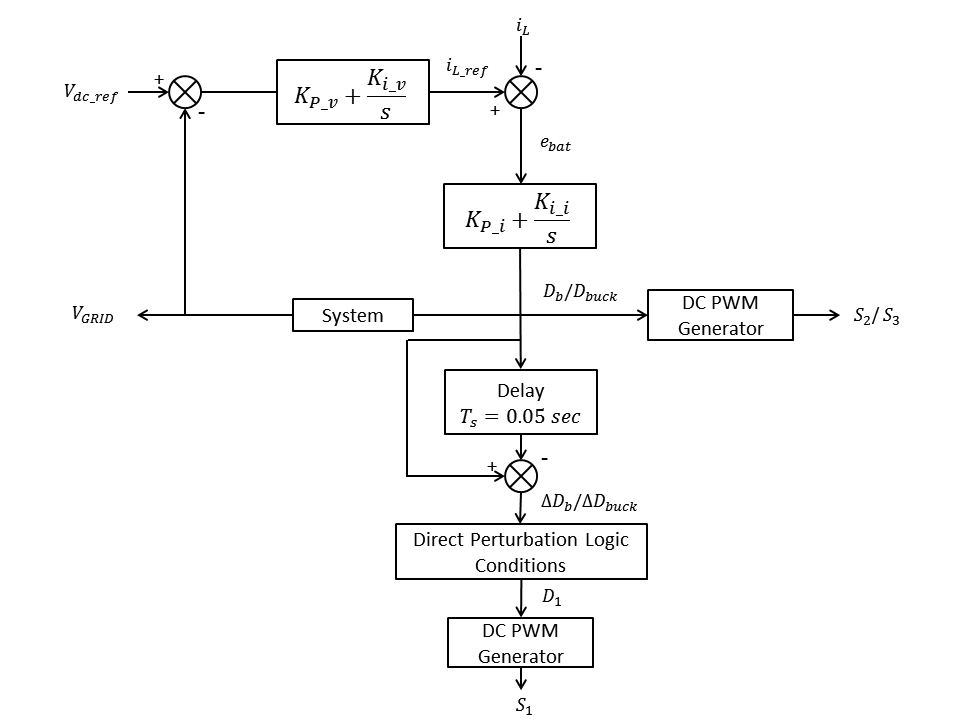}
\caption{Control strategy for MPPT and DC bus voltage control}
\label{fig:blockdiagram}
\end{figure}

The algorithm works in such a novel way that in a single work space not only it tracks MPP but also carries out efficient charging and discharging of a battery by instant transition from buck to boost mode and vice-versa with respect to grid voltage operating conditions for the proposed standalone DC microgrid. A nested control structure is proposed which relates all the control inputs and outputs to operate in an integrated form and is well understood from block diagrams in Fig. \ref{fig:blockdiagram}. Duty ratios calculated in one section of the algorithm are simultaneously used by voltage and current loops to decide the mode of operation for bi-directional converter. 

\subsection{Direct perturbation based sensor-free MPP tracking}
In this approach, a new direct perturbation based sensor-free MPPT algorithm is developed. Unlike the basic P$\&$O method, the current algorithm does not use any sensor measurement at PV array to observe a corresponding change in power; instead the change in power is estimated by observing the grid voltage sensor measurement. On which the decision is appropriately taken automatically which gradually progresses towards MPP; hence the name direct perturbation based is given.
If a perturbation on duty cycle of the boost converter causes an increase in PV power, it will momentarily increase the grid voltage. This change is sensed by a voltage sensor and provides it to the PI controller. The output of PI controller is the duty cycle of buck mode of operation for bi-directional converter which will be decreased in steps to reduce the supplied power from battery so that excess power caused by perturbation can be counter balanced. Therefore a reduction in that duty cycle can be viewed as increase in PV power and hence decision of perturbation can be continued in same direction (considered as positive). Similarly decision to reverse the direction of perturbation (negative) is taken if the corresponding duty cycle increases as an effect to restore the grid voltage. \\

The whole instructions work in a cause-effect relationship. A change in duty ratio of the boost converter $\Delta D_{pv}$ associated directly with the PV source is considered as the cause. Change in duty ratio of either buck or boost mode of bi-directional converter 
( $\Delta D_b^{'}$ or  $\Delta D_{b}$)  is the effect. In every iteration, the sign of the quantities $\Delta D_{pv} \times \Delta D_{b}$ and $\Delta D_{pv}\times\Delta D_b^{'}$ are checked to take decision on next perturbation. The algorithm approaches MPP by continously executing the set of conditions given in Table \ref{tab:buck} and Table \ref{tab:boost}. A small part of the algorithm is illustrated as follows:
\begin{eqnarray}
\Delta D_{pv}= D_{pv}-D_{pv_{old}}; \nonumber \\ \nonumber
\Delta D_{b}= D_{b}-D_{b_{old}} \\ \nonumber
if (mod(t,0.05)==0) \nonumber \\
    if (\Delta D_{pv})*(\Delta D_{b}) > 0; \nonumber\\
    D_{pv_n}=D_{pv}-delD;             \\                                            
    D_{pv_{old}}=D_{pv};\nonumber  \\
    D_{pv}=D_{pv_n};\nonumber\\
    D_{b_{old}}=D_b; \nonumber
\end{eqnarray}

Where $D_{pv}$ and $D_{b}$ are the duty ratio of boost converter and boost mode operation of BDC. The same way computation is done for other set of conditions given in Table \ref{tab:buck} and Table \ref{tab:boost}.

\begin{table}[htbp] 
\caption{Set of conditions for buck operation}
\begin{center}
\begin{tabular}{|c|c|c|c|c|} 
\hline
 \textbf{S.No} & \textbf{Cause}&  \textbf{Effect} & \textbf{Product} & \textbf{Decision}  \\
 & \textbf{$\Delta D_{pv} $}& \textbf{$\Delta D_b^{'} $}& \textbf{$\Delta D_{pv} * \Delta D_b^{'} $} &  $\Delta D_{pv} $ \\
\hline
Case-1 & Positive & Positive & Positive & Positive  \\
\hline
Case-2 & Positive & Negative & Negative & Negative  \\
\hline
Case-3 & Negative & Positive & Negative & Negative  \\
\hline
Case-4 & Negative & Negative & Positive & Positive  \\
\hline
\end{tabular}
\label{tab:buck}
\end{center}
\end{table}

\begin{table}[htbp]
\caption{Set of conditions for boost operation}
\begin{center}
\begin{tabular}{|c|c|c|c|c|} 
\hline
 \textbf{S.No} & \textbf{Cause}&  \textbf{Effect} & \textbf{Product} & \textbf{Decison}  \\
   & \textbf{$\Delta D_{pv} $}& \textbf{$\Delta D_{b} $}& \textbf{$\Delta D_{pv} * \Delta D_{b} $} & Decision $\Delta D_{pv} $ \\
\hline
Case-1 & Positive & Positive & Positive & Negative  \\
\hline
Case-2 & Positive & Negative & Negative & Positive  \\
\hline
Case-3 & Negative & Positive & Negative & Positive  \\
\hline
Case-4 & Negative & Negative & Positive & Negative  \\
\hline
\end{tabular}
\label{tab:boost}
\end{center}
\end{table}

\subsection{Automatic battery charging/discharging with seamless mode transition for DC grid voltage control}

The objective of this section is to control the automatic charging and discharging of the battery system while performing the MPPT. For this a dual loop PI controller is applied constituting an inner current and outer voltage loop. The current loop works with more speed than the voltage loop due to the faster dynamics associated with current. The inner loop checks the direction of inductor current associated with BESS while the outer loop takes care about the DC grid output voltage. Firstly the difference of reference DC ($V_{dc_{ref}}$) and grid voltage ($V_G$) is processed through a simple PI controller to get the reference value for battery inductor current ($i_{l_{ref}}$). Another PI controller associated with the inner loop gives rise to $D_b$ and $D_b^{'}$   from the difference between ($i_{l_{ref}}$)  and ($i_l$). 
This can be expressed as below:
\begin{equation}
 i_{l_{ref}}=(V_{dc_{ref}}(s)-V_{G}(s))\times(K_{p_v}+\frac{K_{i_v}}{s}) 
\end{equation}

\begin{equation}
 D_b = (i_{l_{ref}}(s)-i_l(s)\times(K_{p_i}+\frac{K_{i_i}}{s})     
\end{equation}

\begin{equation}
 D_b^{'} = -(i_{l_{ref}}(s)-i_l(s))\times(K_{p_i}+\frac{K_{i_i}}{s}) 
\end{equation}

where $K_{p_v}$,$K_{i_v}$ and $K_{p_i}$,$K_{i_i}$ are proportional and integral controller gains for voltage and current loops respectively, $D_b$ and $D_b^{'}$ are duty ratios of boost and buck mode of operation respectively for the BDC. Depending on the output of PI controller, the BDC operates in two modes as follows:

\begin{itemize}
    \item When $i_{l_{ref}} < 0$, BDC operates in buck mode and excess power is given back to the storage device to restore DC bus voltage and it gets charged. The algorithm with set of logic conditions for buck operation is tabulated in Table \ref{tab:buck}. 
    \item When $i_{l_{ref}} > 0 $, BDC operates in boost mode and shortage of power is supplied by the storage device to the DC bus and again controlling its voltage to a reference value. The algorithm for boost operation is tabulate in Table \ref{tab:boost}.
\end{itemize}

 The transition from buck to boost mode and vice-versa is done smoothly without any considerable delay so that the system outage is avoided. The whole process can be summed up by following equations:
\begin{eqnarray}
if(mod(t,0.0002)==0) \nonumber \\
      e_{bat}=i_{l_{ref}}-i_l \nonumber \\
      sum=sum+e_{bat} \nonumber \\
      if (i_{l_{ref}}>0)            \\                                         
      D_b=K_{p_i}*e_{bat}+ K_{i_i}*sum; \nonumber \\
      if (i_{l_{ref}}<0) \nonumber \\
     D_b^{'}=  -(K_{p_i}*e_{bat}+K_{i_i}*sum; \nonumber
\end{eqnarray}

\section{Simulation Results}\label{sec:uniresults}
\begin{figure}[!ht]
\centering
\includegraphics[width=0.8\linewidth]{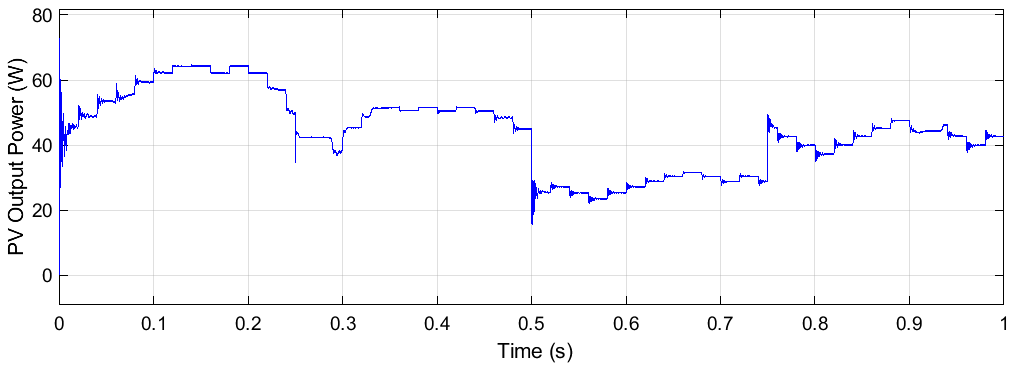}
\caption{PV power varaiation due to change in solar irradiance}
\label{fig:power}
\end{figure}

The complete low voltage DC microgrid circuit is designed and implemented using MATLAB Simulink. The circuit parameter for the simulation is taken from Table \ref{tab1}. The grid reference voltage is set at 20 $V$ and load resistane is taken to be $ 10 \Omega $. Initially solar irradiance is set at 1000 $W/m^2$ . Its value is decreased to 800 $W/m^2$ at 0.25 $s$ of the simulation.The irradiance is further decreased to 500 $W/m^2$ at 0.5 $s$ and then increased to 750 $W/m^2$ at 0.75 $s$. The variation of solar power due to change in solar irradiance is shown in Fig. \ref{fig:power}.

\begin{figure}[!ht]
\centering
\includegraphics[width=0.8\linewidth]{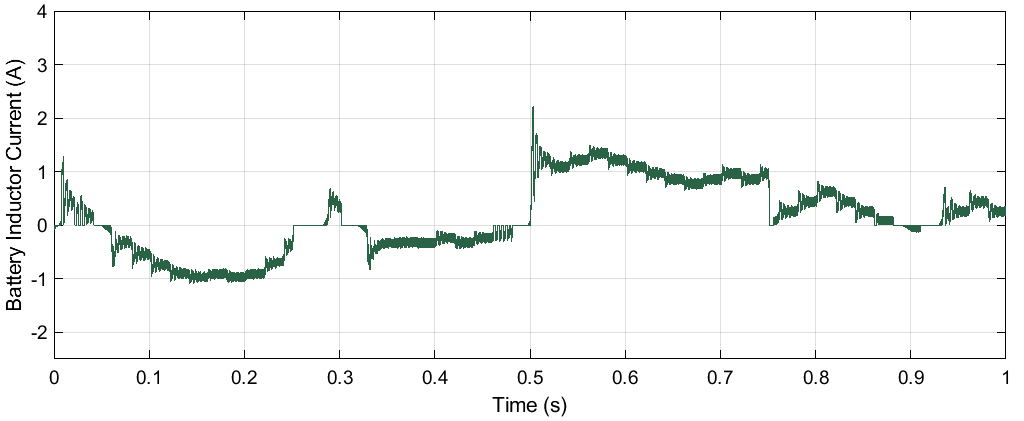}
\caption{Inductor current associated with BESS}
\label{fig:inductor current}
\end{figure}

From 0 $s$. to 0.25 $s$ and from  0.25 $s$ to  0.5 $s$ , BDC works in buck mode indicating that excess of power at DC bus being supplied to the battery to restore 20 Volts at DC bus. In this mode of operation, current direction is negative so that battery gets charged.  This is clearly depicted in Fig. \ref{fig:inductor current} where reference inductor current is below zero up to 0.05 $s$ justifying the accurate control action.

Due to irradiance reduction after $0.5s$ , the voltage at load side tends to change. But the control strategy has to restore the dc bus or load voltage. Therefore BDC has to operate in boost mode so that shortage of PV power at load side is required to supply from battery. In this action the battery discharges and its current starts to flow towards load indicating its reference value above zero.

\begin{figure}[!ht]
\centering
\includegraphics[width=0.8\linewidth]{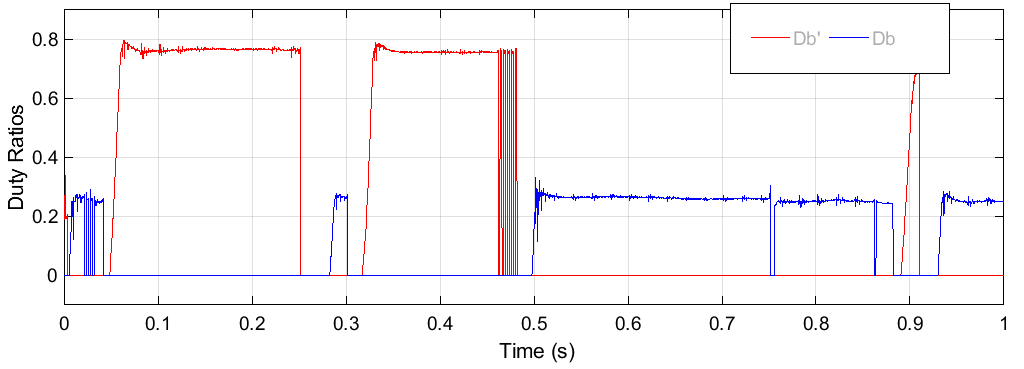}
\caption{Duty ratio of buck and boost operation}
\label{fig:duty ratio}
\end{figure}

The corresponding duty ratios for buck and boost mode of operation respectively is given by $D_b^{'}$ and $D_b$ can be seen in Fig. \ref{fig:duty ratio}. It shows the corresponding buck and boost transition is smoothly done with respect to grid voltage operating condition. It is carried out without any delay to avoid unnecessary outages in the system.

\begin{figure}[!ht]
\centering
\includegraphics[width=0.8\linewidth]{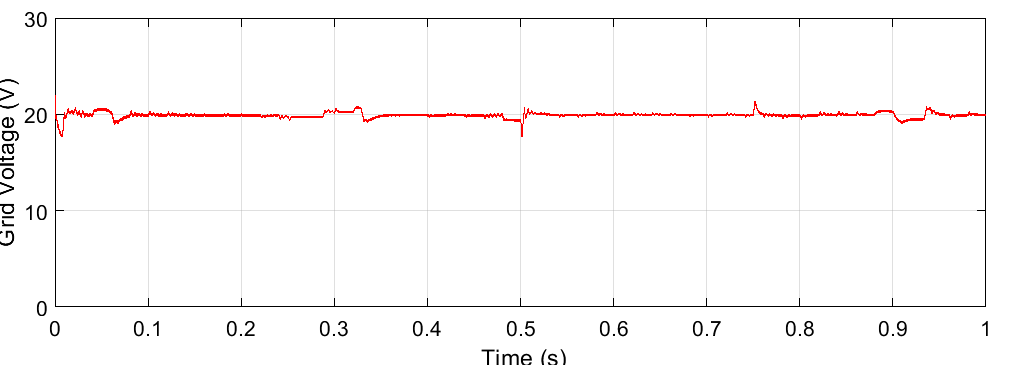}
\caption{DC bus voltage}
\label{fig:bus voltage}
\end{figure}
Fig. \ref{fig:bus voltage} shows that the output voltage at DC bus/grid remains constant at a set reference value so that during any disturbance controller restore the voltage and constant power is supplied to the load.

The results show that the developed closed loop system is extremely stable when disturbances occur in the form of irradiation change. A detailed stability analysis of the closed loop system was carried out after omitting the high speed inner current loop. However, due to lack of space, this has been omitted.

\section{Summary} \label{sec:unisummary}
A direct perturbation based sensor-free MPPT with DC bus voltage control is presented for a standalone DC microgrid unit. In this method MPPT is achieved without using any sensor on PV array side. The controller is able to track the MPPT according to change in the duty ratio of BDC with reference to grid voltage condition. The dual loop PI controller controls the power flow in both the ways by checking the direction of reference inductor current, thus carries out effective battery charging/discharging.  The transition from buck to boost mode is done efficiently without any delay. From simulation results it is clear that the control algorithm not only regulates DC bus/ grid voltage but also operates PV array at MPP at low cost.

%% file: Content/04.tex
\chapter{A Nonlinear Back-stepping based Controller for Standalone PVDG Systems} \label{chapter:nonlinear}\index{Strict feedback}\index{Lyapunov Stability} \index{Nonlinear Controller}
\chaptermark{BACKSTEPPING CONTROLLER DESIGN}
In this chapter, we will focus on developing a control technique for SPVDG system to tackle various large disturbances with the help of back-stepping strategy.
It is interesting to see how a seemingly generalized nonlinear system model has been viewed in a piece-wise strict-feedback format so that back-stepping could be applied satisfactorily. Towards the end, the application of this control technique to various large disturbance is studied and evaluated. 
\section{Introduction}

Managing the power flow with advanced control strategies has become unavoidable in SPVDG systems and are generally sorted out by adopting a hierarchical control structure. The major control challenges for an SPVDG with battery energy storage are maximum power point tracking and DC voltage control. 

The authors in \cite{nl1}, \cite{nl2} propose a model predictive control technique on a boost converter setup.  However, the controller design is linear in nature which makes enables the system to operate only in a limited manner. It is limited in the sense that it does not cater to large changes in operating points. \cite{nl3} describes a passivity type controller for a similar boost converter setup but the system is very sensitive to exact parameter tuning which may create problems during implementation. \cite{nl4} adopts a sliding mode based design to appropriately charge or discharge the battery to maintain voltage. However, lot of precaution needs to be taken to select a sliding surface that can  avoid unnecessary switching during sudden disturbances.

In course of its operation, an SPVDG encounters a large region of operation and thus a far advanced and complex control strategy is justified. Backstepping controller is a good choice of controller design as it operates in a larger range of operation  especially when the system is underactuated like in the SPVDG case. Moreover, a detailed stability analysis can be carried out with this kind of controllers which is very much desirable for these type of systems. A backstepping type control has been designed in a shipboard power system to control voltage in \cite{nl5}. \cite{nl6,nl7} applied similar back-stepping technique in microgrids for voltage control. 

However, all these works have been developed to target only a specific type of control in their overall system. This chapter develops a generic modular back-stepping based design that can be applied for achieving multiple control goals in the SPVDG system. Also, a plug and play approach as the one developed in this work, allows independent operation of each device in system assuring greater flexibility while improving reliability. Finally, the use of such nonlinear techniques also improves the speed of response in case of transients. It is also very easy to scale up to higher order systems with many sub-systems. The system’s stability is shown by constructing a series of Lyapunov functions. The overall system then guarantees good dynamic performance along with flexibility.

 The following is the organization of this chapter. Section \ref{sec:nlmodel} describes the SPVDG system under consideration, its mathematical modeling and its control hierarchy. Section \ref{sec:nlcontrol} gives the complete stability proof of the piece-wise back-stepping control along with the designed controllers and observers. Section \ref{sec:nlresults} shows the results that are obtained due to occurrence of different intermittencies while Section \ref{sec:nlsummary} concludes the chapter. 
 
\section{System Description and Modeling} \label{sec:nlmodel}

This section explains the details of different building blocks of the SPVDG system, its control structure and the large signal modeling of the overall system.
\subsection{Basic Blocks} 
\begin{figure}[!t]
\centering
\includegraphics[width=0.8\linewidth]{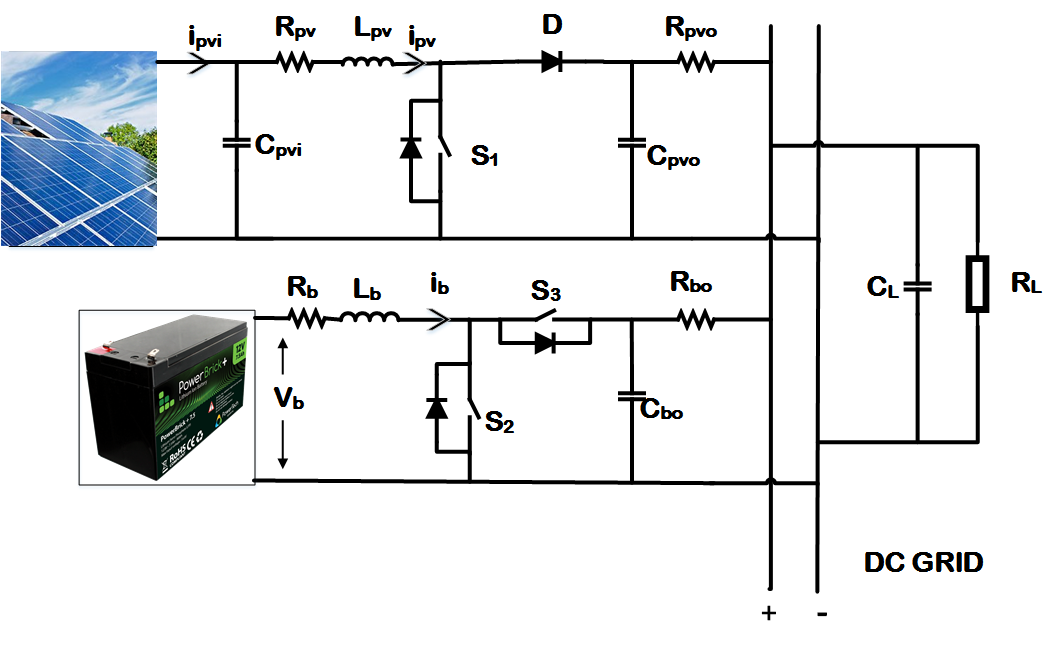}
\caption{Power Architecture of the SPVDG System}
\label{fig:spvdg2}
\end{figure}

The SPVDG system under consideration consists of a PV array and a battery energy storage system (BESS) feeding to a load. The PV array continuously extracts power from the available solar energy and transfers it to the load using a DC-DC converter as shown in Fig. \ref{fig:spvdg2}. The BESS is connected to the grid via a bidirectional DC-DC converter. This is essential to modulate power flow in both the directions as demanded by load conditions in the SPVDG system. Both the converters are controlled using to regulate the desired flow of power in the PVDG system. 

\begin{figure}[!b]
\centering
\includegraphics[width=0.8\linewidth]{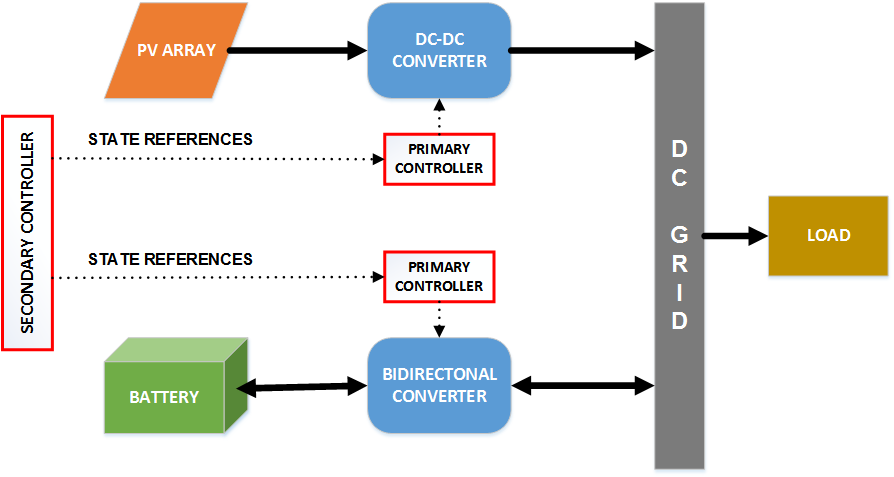}
\caption{Structure of SPVDG System}
\label{fig:spvdg0}
\end{figure}

\subsection{Hierarchical Control Structure} \index{Hierarchical Control}
 As evident from Fig.\ref{fig:spvdg0}, the hierarchy of control in the SPVDG system is present in two levels of the SPVDG system  namely, primary and secondary. The secondary level controller handles the overall energy management function of the SPVDG. It is responsible for ensuring maximum power extraction from the PV panel and also plan the power flow of the battery depending on the imbalance between the PV power generation and load consumption. The secondary controller achieves this operation by setting the reference values for all the important states in the SPVDG system like inductor currents and capacitor voltages. The major function of the primary controller is to bring the system states to the reference values set by the secondary controller. A properly designed primary controller performs this function even in the presence of large disturbances affecting the system time to time.

 \subsection{State Space Model} \index{SSA! State Space Averaging}
The power circuit architecture of SPVDG system is shown in Fig.\ref{fig:spvdg2} which is mathematically modelled in this section. The well established State Space Averaging \cite{ssa} technique is used for modelling the overall system due to abundance of power electronics in the system.  

The PV system consists of a PV array whose output current equation is given by \eqref{eq:ipv}
\begin{align}\label{eq:ipv}
\scalebox{.9}{$
i_{pv}=n_{p}I_{g}-n_{p}I_{s}\left(e^{\dfrac{q(v_{pv}+i_{pv}R_{s})}{n_{s}pKT}}-1\right)-\dfrac{v_{pv}+i_{pv}R_{s}}{R_{sh}}$}
\end{align}

The PV output current $i_{pv}$ varies with change in temperature and irradiance and it needs to be continuously monitored. In the state-space model, this current has been considered as disturbance $d_1$. The output voltage of PV panel which is also the voltage across capacitor $C_{pvi}$ has been designated as $x_1$, the voltage across capacitor $C_{pvo}$ has been termed as $x_2$ and the current through inductor  $L_{pv}$ has been designated $x_3$. All these states belong to the PV subsystem. 

The battery has been modelled as a voltage source whose voltage is considered as disturbance $d_2$. Similarly, the current through inductor $L_b$ has been termed as $x_4$ and voltage across capacitor $C_{bo}$ has been designated as $x_5$. Both these states belong to the battery subsystem. Finally, the DC grid voltage which is the voltage across $C_L$ and is taken to be $x_6$ and  load admittance $\frac{1}{R_L}$ is modeled as disturbance $d_3$. 
 \newpage
The state space model derived for this system is obtained as follows:
 \begin{eqnarray}
 \dot{x_1}&=& \frac{d_1}{C_{pvi}} - \frac{x_3}{C_{pvi}} \nonumber \\
 \dot{x_2}&=& \frac{x_3}{C_{pvo}} - \frac{x_2}{R_{pvo}C_{pvo}} + \frac{x_6}{R_{pvo}C_{pvo}}- \frac{x_{3}}{C_{pvo}}u_{1} \nonumber \\
 \dot{x_3}&=& \frac{x_1}{L_{pv}} - \frac{x_2}{L_{pv}} - \frac{x_{3}R_{pv}}{L_{pv}} + \frac{x_2}{L_{pv}}u_1 \nonumber \\  
  \dot{x_4}&=& \frac{d_2}{L_{b}} - \frac{x_{4}R_{b}}{L_{b}} - \frac{x_5}{L_{b}} + \frac{x_5}{L_{b}}u_2  \label{eq:ssmodel}\\  
 \dot{x_5}&=& \frac{x_4}{C_{bo}} - \frac{x_5}{R_{bo}C_{bo}} + \frac{x_6}{R_{bo}C_{bo}}- \frac{x_{4}}{C_{bo}}u_{2} \nonumber \\
 \dot{x_6}&=& \frac{x_2}{C_{L}R_{pvo}} +\frac{x_5}{C_{L}R_{bo}}-\frac{x_6}{C_L}\Bigg[\frac{1}{R_{pvo}}+ \frac{1}{R_{bo}} + d_3 \Bigg] \nonumber 
 \end{eqnarray}
 where
 \begin{eqnarray}
  x &=& [V_{pvi}~~V_{pvo}~~i_{pv}~~i_b~~V_{bo}~~V_L ] \nonumber \\
 d_1 &=& i_{pvi} \hspace{2cm} d_2= V_b  \hspace{2cm} d_3= \frac{1}{R_L} \nonumber
 \end{eqnarray}

\section{Reference Generation:} \index{Secondary Control} 
This section shows a technique using which the secondary level references can be generated. The secondary level references are generated by the secondary controller and sent to the primary level controllers so that appropriate control action is taken to follow these secondary level references. Successful calculation/estimation of these references holds the key to power balance of the entire SPVDG system. It also decides whether the bidirectional converter should work in the charging mode or discharging mode. 

Since, the state space model of the system is known, by equating the state space equations to zero and solving the equations by eliminating the control variables is the way forward for generating these references. However, it is to be understood that this would not bring the system to work at our desired voltage and current levels. Hence, we need to substitute certain desired voltage and current values in the previous set of equations obtained through equating the state space equations to zero. Then, we will be able to arrive at the set of desired values  for all the states in the system. 

Let us say that the microgrid needs to operate at a voltage $V_{DC}$ and that the PV output voltage should be maintained at $V_{mpp}$ to ensure maximum power extraction from the panel/array. These two references serve the basis for generating the references of all the states in the system. The following set of equations denote the values of references used for the current SPVDG system:
\begin{eqnarray}
x_{6ref} &=& V_{DC}(known) \nonumber \\
x_{1ref}&=& V_{mpp}(known) \nonumber\\
x_{3ref}&=& I_{mpp}(known)\nonumber \\
& &a_1=1 \hspace{1cm}  b_1=-x_{6ref} \nonumber \\
& &c_1= x_{3ref}^2R_{pvo}R_{pv}-x_{1ref}x_{3ref}R_{pvo}\nonumber \\
x_{2ref}&=&-b_1+ \sqrt{\frac{b^2_1-4a_1c_1}{2a_1}} \nonumber \\
x_{5ref}&=&-\frac{R_{bo}}{R_{pvo}}x_{2ref} + x_{6ref}\Bigg(1+\frac{R_{bo}}{R_{pvo}}+\frac{R_{bo}}{R_l}\Bigg) \nonumber \\ 
& & a_2= R_{bo}  \hspace{1cm}  b_2= -R_{bo}d_2 \nonumber \\
& &c_2= x_{5ref}^2-x_{5ref}x_{6ref} \nonumber \\ 
x_{4ref}&=&-b_2+ \sqrt{\frac{b^2_2-4a_2c_2}{2a_2}}  \nonumber \\
\end{eqnarray}

Although, all the references are generated initially using this technique, in the actual implementation, a combination of references obtained from this section and the virtual control references generated in the upcoming section will be utilized for maximizing the effect of back-stepping control.
 
\section{Backstepping based Nonlinear Controller Design} \label{sec:nlcontrol} \index{Back-stepping}
This section delineates the detailed procedure for designing the back-stepping based controller proposed in this chapter. First,  the complete mathematical design of the different controllers present in the SPVDG system is explained which is then followed by a summary of the entire procedure enumerated point-wise. The mathematical derivation of the controllers also serves as the complete stability analysis of the entire design technique. Hence, it serves a two-way purpose.
 
\subsection{Controller Design Procedure:}
    
Let us assume a Lyapunov function $V_1=\frac{C_{pvi}}{2}e^2_1$ where $e_1= x_1-x_{1ref}$. Upon differentiation, we get $\dot{V}_1= e_1(d_1-x_3)$. If we choose $x_3$ as
\begin{eqnarray}
\alpha_3 = d_1 + K_1e_1 \label{eq:nvu3}
\end{eqnarray} 
then ,$\dot{V}_1= -K_1e^2_1$ which means $\dot{V}_1<0$. Then, we consider the Lyapunov function $V_{2,3}= \frac{C_{pvo}}{2}e^2_2 + \frac{L_{pv}}{2}e^2_3$ where, $e_2= x_2-x_{2ref}$ and $e_3= x_3-\alpha_3$ .  Differentiating we get,

\begin{eqnarray}
\dot{V}_{2,3}&=& C_{pvo}e_2\dot{x}_2 + L_{pv}e_3(\dot{x}_3 - \dot{\alpha}_3) \nonumber \\
&=& e_2\Bigg(x_3 + \frac{x_6-x_2}{R_{pvo}} \Bigg) + e_3(x_1-x_2-R_{pv}x_3-L_{pv}\dot{\alpha}_3) \nonumber \\
& & + u_1(-e_2x_3 + e_3x_2) \nonumber \\
\text{If $u_1$ is chosen as follows:} \nonumber \\
u_1 &=& \frac{(-num_{c1} - K_2{e_2^2} - K_3{e_3^2})}{e_3x_2- e_2x_3} \label{eq:nu1}\\
\text{where} \nonumber\\
num_{c1}&=& e_2x_3 - \frac{e_2x_2}{R_{pvo}} + \frac{e_2x_6}{R_{pvo}} + e_3x_1 - e_3x_2 - e_3x_3R_{pv} - e_3L_{pv}\dot{\alpha}_3  \nonumber  \\ \\
\dot{\alpha}_3&=& K_1 \dot{x}_1 + \dot{d}_1 
\end{eqnarray}
then, it results in $\dot{V}_{2,3}= -K_2e^2_2 -K_3e^2_3$ which means $\dot{V}_{2,3} < 0$.
Using this value of $u_1$, the convergence of states $x_1$, $x_2$ and $x_3$ of the PV array is ensured.
  
Now, for finding the next controller, first we choose a Lyapunov function \\ $V_6=\frac{C_L}{2}e^2_6$ where $e_6= x_6 - x_{6ref}$ .

\begin{eqnarray}
\dot{V}_6 &=& C_L e_6 \dot{x}_6  \nonumber \\
&=& e_6\Bigg[\frac{x_2}{R_{pvo}}+\frac{x_5}{R_{bo}}- x_6\Bigg(\frac{1}{R_{pvo}}+\frac{1}{R_{bo}}+d_3 \Bigg)\Bigg] \nonumber
\end{eqnarray}
If, the value of $x_5$ equals reference virtual input $\alpha_5$  where 
\begin{eqnarray}
\alpha_5= -\frac{R_{bo}}{R_{pvo}}x_2 + R_{bo}x
_6\Bigg(\frac{1}{R_{pvo}}+\frac{1}{R_{bo}}+d_3 \Bigg) - K_6e_6R_{bo} \label{eq:nvu5}
\end{eqnarray}
It then becomes $\dot{V}_6=-K_6e^2_6$. Following this step, we choose $V_{4,5}=\frac{C_{bo}}{2}e^2_5 + \frac{L_b}{2}e^2_4$ where $e_5= x_5 - \alpha_5$ and $e_4=x_4-x_{4ref}$. Upon differentiation we get,

\begin{eqnarray}
\dot{V}_{4,5}&=&e_5\Bigg(x_4 - \frac{x_5}{R_{bo}} + \frac{x_4}{R_{bo}}-x_4u_2 - C_{bo}\dot{\alpha}_5\Bigg) + e_4(d_2-x_5-R_bx_4+x_5u_2) \nonumber \\
\end{eqnarray}

We choose $u_2$ as follows to stabilize the system:
\begin{eqnarray}
u_2&=& \frac{-num_{c2}-K_4e^2_4- K_5e^2_5}{e_4x_5-e_5x_4} \label{eq:nu2} \\ 
\text{where} \nonumber \\
num_{c2}&=& e_5\Bigg(x_4+ \frac{x_6-x_5}{R_{bo}} - C_{bo}\dot{\alpha}_5 \Bigg) +e_4(d_2 -x_5-R_bx_4) \\
\dot{\alpha}_5&=& -\frac{R_{bo}}{R_{pvo}}\dot{x}_2 + R_{bo}x_6\dot{d}_3 + \dot{x}_6\Bigg(-K_6R_{bo}+\frac{R_{bo}}{R_{pvo}}+1+R_{bo}d_3\Bigg)
\end{eqnarray}

Then, we get, $\dot{V}_{4,5}= -K_4 e_4^2 - K_5 e_5^2 $  which is equivalent to 
$~~~\dot{V}_{4,5}<0$ for $K_4,K_5>0$.
 
Thus the total Lyapunov function,
\begin{eqnarray}
\dot{V}= \dot{V}_1 + \dot{V}_{2,3} + \dot{V}_{4,5} + \dot{V}_6 < 0
\end{eqnarray}

Thus, adopting the controllers designed in \eqref{eq:nu1}, \eqref{eq:nu2} and the virtual controllers designed in \eqref{eq:nvu3} and \eqref{eq:nvu5} which are carefully derived on the basis of Lyapunov Stability Theory, the stability of all the system states is guaranteed.

The entire procedure for designing both the MPPT and DC bus voltage controllers can be summarized in the following set of instructions. 

\begin{enumerate}
\item Compute the reference values for all states.
\item Compute virtual input $\alpha_3$ for the PV system to make $x_1$ follow $x_{1ref}$.
\item Compute the value of $u_1$ so that $x_3$ follows $\alpha_3$ and $x_2$ follows $x_{2ref}$.
\item Compute the value of virutal input $\alpha_5$ such that $x_6$ reaches its reference value $x_{6ref}$.
\item Compute the value of $u_2$ such that $x_5$ follows $\alpha_5$ and $x_4$ follows its reference value $x_{4ref}$.
\end{enumerate}

\section{Results} \label{sec:nlresults}
In this section, different scenarios have been considered where the effects of change in irradiation, temperature and load were studied when the proposed back-stepping control is applied. The DC bus voltage is taken to be $40$ V. The PV array always operates at a peak power rating of $200$W. Tha battery operates at a constant operating voltage of 24$V$. The simulation has been carried out in MATLAB Table-\ref{tab0} shows gives an overall picture of various system and controller parameters used for the simulation.
\begin{table}[htbp] \label{table:parameters}
\caption{Parameter Specification}
\begin{center}
\begin{tabular}{|c|c|}
\hline
\textbf{Subsystem} & \textbf{\textit{Parameter Specification}} \\
\hline
Battery Voltage &  $24 V$ \\
\hline
PV Array at STC & Model: Kyocera Solar KC200GT \\ 
& $V_{OC}=32.9V$, $I_{SC}=8.21A$ \\
& $V_{MPP}=26.3V$, $I_{MPP}=7.61A$ \\  
\hline
DC/DC Converter & $C_{pvi}=3mF$, $C_{pvo}=3mF $, $L_{pv}=10 mH$, \\   & $R_{pv}=0.5\Omega$ , $R_{pvo}=0.1\Omega$ \\
\hline
DC/DC Bidirectional & $C_{bo}=3mF$, $R_{b}=0.5\Omega $, \\
Converter & $L_{b}=10 mH$, $R_{bo}=0.1\Omega $ \\
\hline
Back-stepping & $K_1$=10, $K_2$=0.04, $K_3$=0.04\\
Gains & $K_4$=0.04, $K_5$=0.04, $K_6$=15.0\\
\hline
\end{tabular}
\label{tab0}
\end{center}
\end{table}

\subsection{Case-1: Variation in Irradiance }
\begin{figure}[h]
	\centering
		\includegraphics[width=\linewidth]{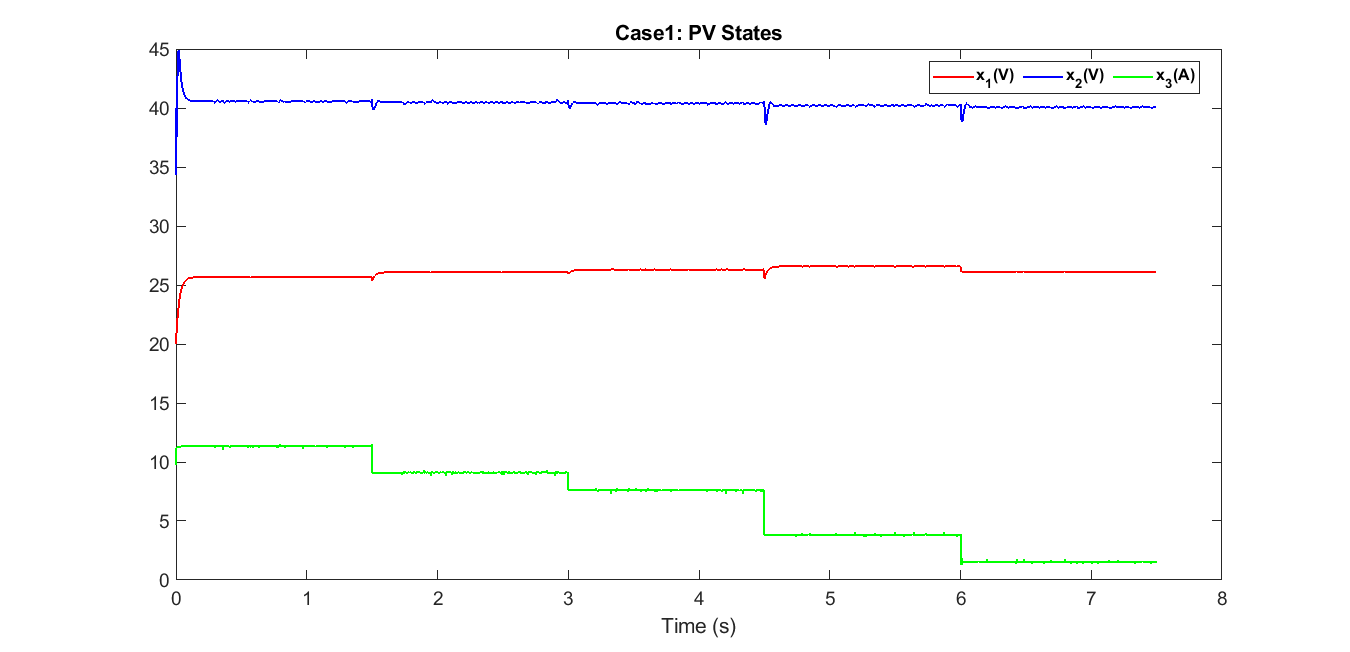}
		\caption{Case-1: PV states}
		\label{fig:case41}
\end{figure}

\begin{figure}[h]
		\centering
		\includegraphics[width=\linewidth]{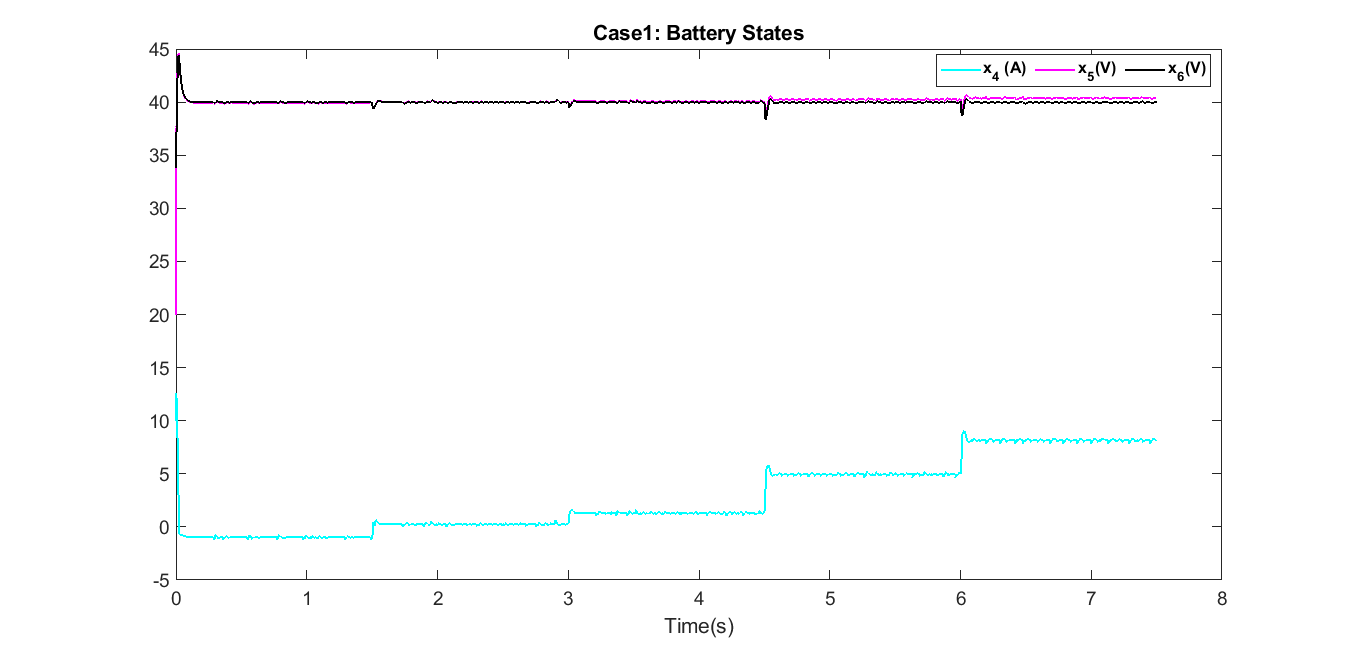}
		\caption{Case-1: Battery States }
		\label{fig:case42}
\end{figure}

In this case, the irradiance of PV panel is changed with time while temperature and load are kept constant at 25$\degree$C and 200$W$. The irradiance changes for every 1.5 second from 1500$W/m^2$ to 1200$W/^2$ and then to 1000$W/m^2$, 500$W/m^2$ and 200$W/m^2$ at 1.5$s$, 3$s$, 4.5$s$ and 6$s$ respectively. Figures \ref{fig:case41} and \ref{fig:case42} show the evolution of PV and battery states for this case.

When irradiance changes, both the voltage and current at which maximum power is extracted from the PV panel change. Moreover, the maximum power that can be extracted also reduces with reduction in irradiance. This effect is visible in the PV and battery currents shown in Fig.\ref{fig:case41}. Since load is constant at 200$W$, as the irradiance decreases, the PV inductor current reduces and the battery current increases compensating the reduction in PV current.  During change in irradiance, some transients occur in DC bus voltage but they die out within 80$ms$. Therefore, the DC bus voltage remains constant at 40$V$.

\subsection{Case-2: Variation in Temperature}
With increase in temperature, the maximum power extracted from the PV panel reduces, and also the values of MPP current and voltage. In this case, the temperature is varied keeping load and irradiation constant at 200$W$, 1000$W/m^2$ respectively. The temperature is varied from 75$\degree$ C to 50$\degree$ C and to 25$\degree$ C, 10$\degree$ C and 0$\degree$ C respectively, at 1.5$s$, 3$s$, 4.5$s$ and 6$s$ respectively.
\begin{figure}[!h]
	\centering
		\includegraphics[width=\linewidth]{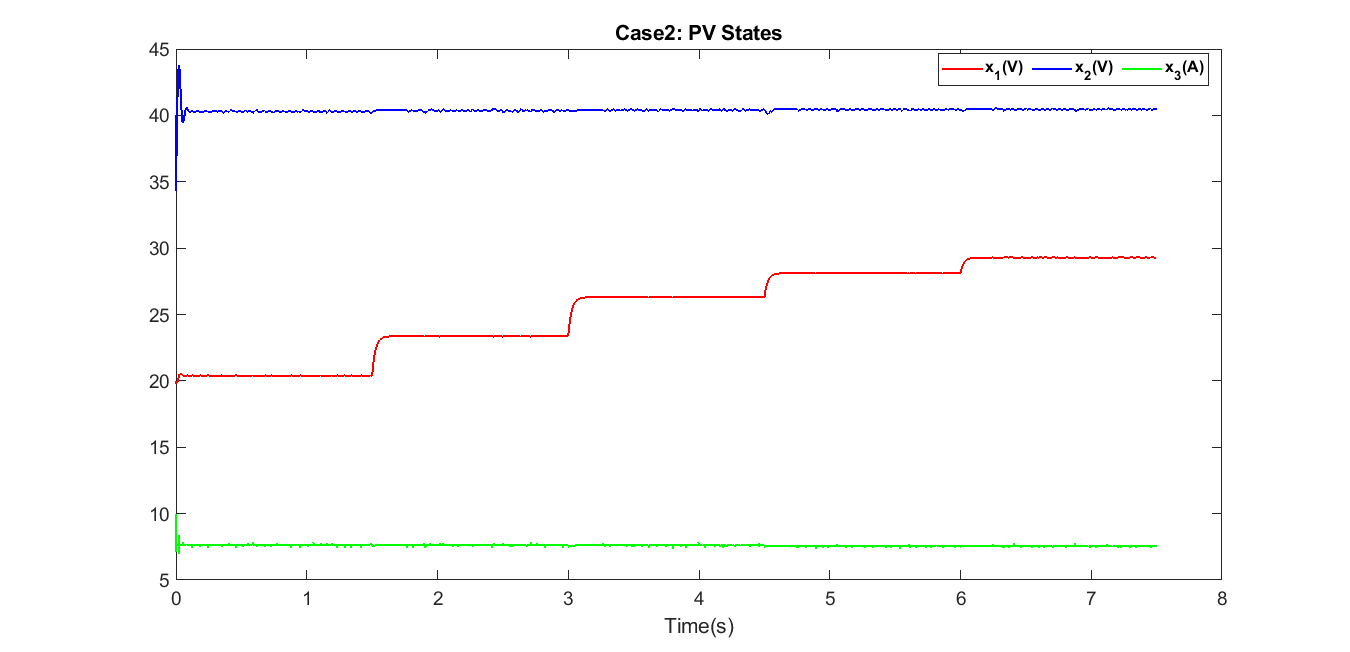}
		\caption{Case-2: PV states}
		\label{fig:case51}
\end{figure}

\begin{figure}[!h]
		\centering
		\includegraphics[width=\linewidth]{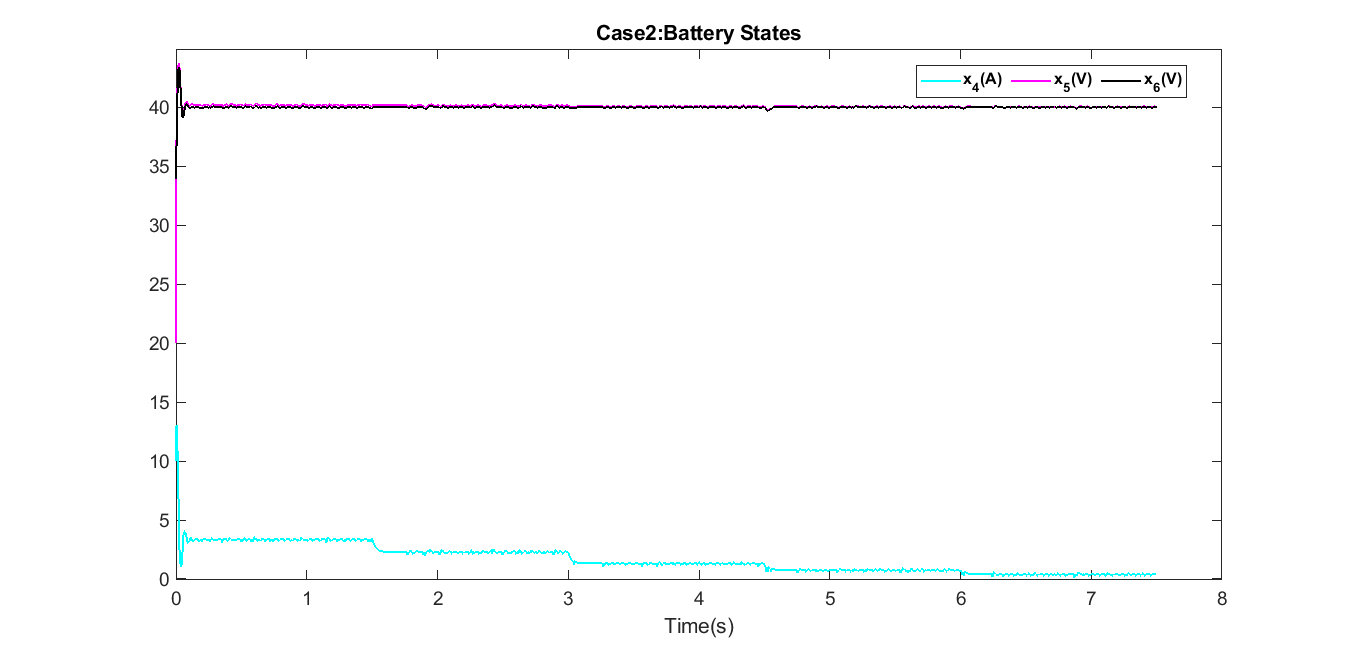}
		\caption{Case-2: Battery States }
		\label{fig:case52}
\end{figure}
Figures \ref{fig:case51} and \ref{fig:case52} show the response of the SPVDG system to change in temperature. However, the variation in MPP current extracted from PV with change in temperature is less compared to the change in MPP voltage. The grid voltage stays at 40$V$ irrespective of any variation in temperature.  

\subsection{Case-3: Variation in Load}
In this case, the PV panel characteristics remain constant with constant temperature and irradiation at 1000$W/m^2$ and 25 $\degree$C. However, the load resistance is varied from 5$\Omega$ to 7$\Omega$, 9$\Omega$, 11$\Omega$ and then to 8$\Omega$ at 1.5$s$, 3$s$, 4.5$s$ and 6$s$ respectively.
\begin{figure}[h]
	\centering
		\includegraphics[width=\linewidth]{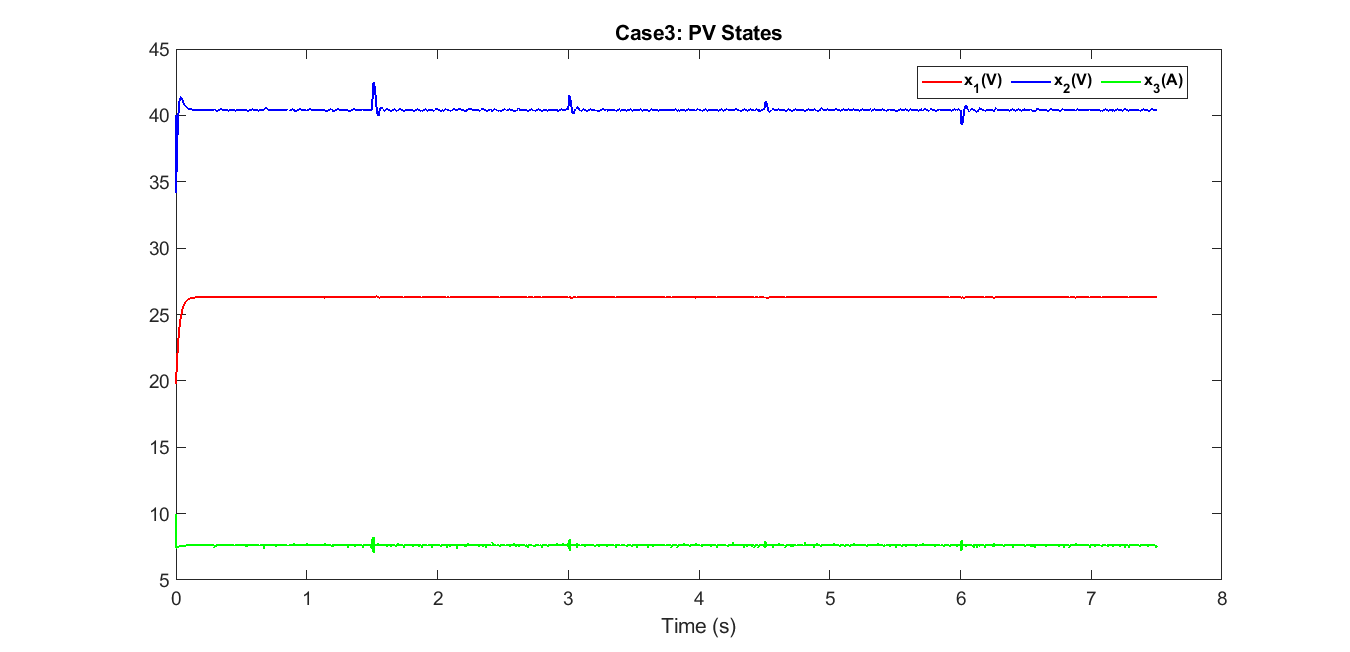}
		\caption{Case-3: PV states}
		\label{fig:case61}
\end{figure}

\begin{figure}[h]
		\centering
		\includegraphics[width=\linewidth]{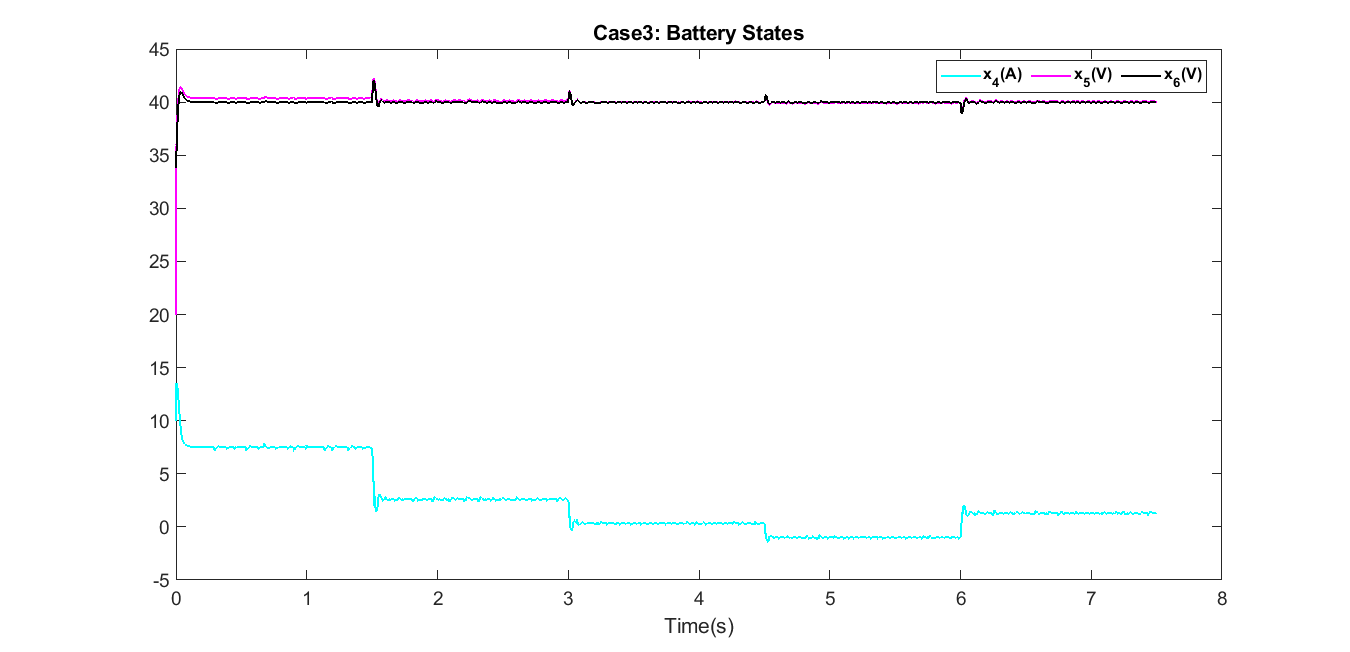}
		\caption{Case-3: Battery States }
		\label{fig:case62}
\end{figure}

The change in battery current can easily be observed when PV generation is constant and load is changed. Initially, when the load is very high, the battery also supplies positive current to assuage the load requirement and gradually, as the load is reduced the current supplied by the battery reduces and when the load reduces below the maximum power provided by the PV panel, the battery current becomes negative showing that the extra power produced by the PV panel is being used to charge the battery. Its worth noting that both the MPP voltage $x_1$ and grid voltage $x_6$ remain constant. 

In all the cases it can be observed that the grid is hardly perturbed without any deviations from desired value. All the states completely stabilize within 80$ms$.  Hence, the proposed algorithm is fast and very robust to change in disturbances.

\section{Summary} \label{sec:nlsummary}
In this chapter, a nonlinear back-stepping control strategy is developed for a standalone PVDG system with battery energy storage. This technique ensures faster stabilization of all the system states and grid voltage when subjected to large disturbances. It has been verified for three different cases in presence of large variations in irradiance, temperature and load. This controller also ensures appropriate bidirectional power flow depending on the power balance in the standalone system.The simulation results validate the efficacy of the proposed nonlinear control strategy. 

%% file: Content/05.tex
\chapter{Disturbance Observer based Backstepping Controller for the Isolated DCMG Unit} \label{chapter:distobs}\index{Disturbance Observer!Back-stepping Design}
\chaptermark{DISTURBANCE OBSERVER BASED BACK-STEPPING DESIGN}

\section{Introduction} 
In this work, the back-stepping strategy developed in the previous chapter is further equipped with disturbance observers in order to reduce the placement of excessive sensors that is demanded by the non-linear control techniques. Lyapunov stability theory is exploited to design the observers which will be clearly delineated in this chapter. 
\begin{figure}[!b]
\centering
\includegraphics[width=0.8\linewidth]{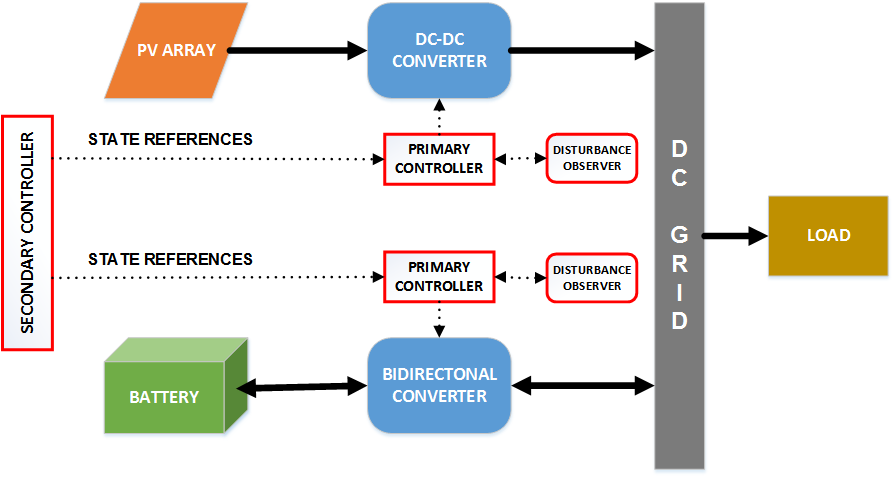}
\caption{Structure of SPVDG System with observer}
\label{fig:spvdg1}
\end{figure}

The variations in operating points during general course of operation are large due to the intermittencies in DG supply and loading conditions\cite{iovine1}. Therefore, it would be very difficult for these controllers to maintain stability during drastic changes like reduction in PV generation, large fluctuations in load, etc. Ensuring stability of nonlinear system models requires complex controllers to be implemented. The increase in complexity of controllers is the trade off for assuring stability inside a whole operation region.

For instance, feedback linearization, sliding mode control and adaptive control based techniques are seen to have been successfully implemented on this system in literature \cite{fbl}\cite{smc}. Nevertheless, all these techniques are model-based which require continuous feedback from a lot of states and parameters existing in the system. This calls for procuring a great number of sensors for measuring these quantities for implementing them successfully. This invariably results in increasing the overall cost of the entire system. However, with the power of modern computing devices it is possible to estimate many of such parameters to feed into the nonlinear control law directly.

Works such as \cite{reducedsensor1} contributed towards reducing sensors for water pumping system  but these do not address MPPT or voltage control in PV based systems. A very simple technique for reducing sensors for a PV battery system was proposed in \cite{isgtpaper} which exploits the power balance of the electrical system to reduce the number of sensors in a DC microgrid but the performance of this controller was traded for making the technique cost effective. This technique will not work for large change in disturbances. 

Hence, in this chapter we propose the following:

\begin{itemize}
    \item  An intuitive plug and play back-stepping based controller design for various subsystems in an islanded PV system with storage.
    \item An update law for observing/estimating various disturbances in the system like output current of the PV array, battery voltage and load in an online fashion.
\end{itemize}

It is also to be noted in this work that the reference values from secondary controller are readily available and hence, will mainly deal with the development of the nonlinear primary level controllers. The observers act on the same level of primary controllers and work on the the basis of states' observations received from different sensors. This can be seen in Fig. \ref{fig:spvdg1}.

The following is the organization of this chapter. Section \ref{sec:controldo} gives the complete stability proof of the disturbance observer based back-stepping control along with the designed controllers and observers. Section \ref{sec:resultsdo} shows the results that are obtained due to occurrence of different intermittencies while Section \ref{sec:summarydo} concludes the chapter.

\section{Disturbance Observer Based Back-stepping Controller Design} \label{sec:controldo}
A cursory glance over the system model in \eqref{eq:ssmodel} reveals that it is highly under-actuated  for which back-stepping based control would be a natural choice to adopt. However, the proposed controller design technique deviates from the original process of back-stepping design since the system model doesn't exactly fit into the well-known strict-feedback structure. Moreover, it is worth noting that the disturbance observers synthesized concomitantly with the control laws were essential in the current context since it was assumed that all the disturbances are unknown. In this section, the complete process of designing the disturbance observer based back-stepping control along with its many nuances is delineated which in itself would also establish the stability of both the controllers and observers designed for the system.

\hspace{-0.3cm}Let us assume a Lyapunov function
 \begin{eqnarray}
 W_1= \frac{C_{pvi}}{2} e_1^2  +  \frac{1}{2\rho_1}\widetilde{d}_1^2
 \end{eqnarray}
where $e_1 = x_1-x_1^d$ .
 Taking the time derivative of $W_1$ we get, 
 \begin{eqnarray}
 \dot{W}_1 &=& C_{pvi}e_1\dot{e}_1  + \frac{1}{\rho_1}\widetilde{d}_1(-\dot{\widehat{d}}_1) \nonumber \\ 
  &=& e_1(d_1 - x_3) - \frac{1}{\rho_1}\widetilde{d}_1(\dot{\widehat{d}}_1) \nonumber \\ &= & e_1(\widetilde{d}_1 +\widehat{d}_1 - x_3) - \frac{1}{\rho_1}\widetilde{d}_1(\dot{\widehat{d}}_1) \nonumber \\
  &=& e_1(\widehat{d}_1 - x_3) + \widetilde{d}_1(e_1 - \frac{\dot{\widehat{d}}_1}{\rho_1})
 \end{eqnarray}
 
 \hspace{-0.3cm}We choose the values of $x_3$ to be equal to $\alpha_3$ and 
$\dot{\widehat{d}}_1$ as follows:
\begin{eqnarray}
\alpha_3 &=& \widehat{d}_1 + K_1 e_1 \nonumber \\
\dot{\widehat{d}}_1 &=& \rho_1 e_1  \label{eq:d1hatdot}
\end{eqnarray}

\hspace{-0.3cm}Substituting these values, we get
$\dot{W}_1 = - K_1 e_1^2$. Assuming $K_1>0$, it can be concluded that $\dot{W}_1 < 0$

\hspace{-0.3cm}Now considering $e_2 = x_2 - x_2^d$ and $e_3=  x_3- \alpha_3$, we define the following Lyapunov function jointly for the second and third states:
\begin{eqnarray}
W_{2,3} &=& \frac{C_{pvo}}{2}e_2^2 + \frac{L_{pv}}{2}e_3^2 \nonumber \\
\dot{W}_{2,3} &=& C_{pvo}e_2 \dot{e}_2 + L_{pv}e_3(\dot{x}_3 - \dot{\alpha}_3)
\end{eqnarray}

\hspace{-0.3cm}Upon rearranging different terms, $\dot{W}_{2,3}$ becomes as follows:
\begin{eqnarray}
\dot{W}_{2,3} &=& e_2(x_3 + \frac{x_6-x_2}{R_{pvo}}) + e_3(x_1 -x_2 - R_{pv}x_3) \nonumber \\ & & + ~ u_1(e_3 x_2 - e_2 x_3) \nonumber
\end{eqnarray}
 
\hspace{-0.3cm}We choose $u_1$ as follows:
\begin{eqnarray}
num_{c1}&=& e_2(x_3 + \frac{x_6-x_2}{R_{pvo}}) + e_3(x_1 -x_2 - R_{pv}x_3) \nonumber \\
den_{c1} &=& e_3 x_2 - e_2 x_3 \nonumber \\
u_1 &=& \frac{-K_2 e_2^2 -K_3 e_3^2 - num_{c1}}{den_{c1}} \label{eq:u1}
\end{eqnarray} 

\hspace{-0.3cm}Thus, we get $\dot{W}_{2,3}= -K_2 e_2^2 - K_3 e_3^2 $  which is equivalent to 
$\dot{W}_{2,3}<0$ for $K_2,K_3>0$. 

\hspace{-0.3cm}The next Lyapunov function is assumed as follows:
\begin{eqnarray}
W_{6}= \frac{C_L}{2}e_6^2 + \frac{1}{2\rho_3}\widetilde{d}_3^2
\end{eqnarray}
Differentiating this function we get,
\begin{eqnarray}
\dot{W}_6&=& e_6\Bigg[\frac{x_2}{R_{pvo}}+\frac{x_5}{R_{bo}}-x_6\Bigg(\frac{1}{R_{pvo}}+\frac{1}{R_{bo}}+\widehat{d}_3\Bigg) \Bigg] \nonumber \\ & & - \widetilde{d}_3\Bigg[\frac{\dot{\widehat{d}}_3}{\rho_3}+ x_6 e_6 \Bigg]
\end{eqnarray}
Choosing $x_5$ to be $\alpha_5$ and  $\dot{\widehat{d}}_3$ as follows:
\begin{eqnarray}
\alpha_5 &=& -R_{bo}K_6 e_6 + R_{bo}x_6\Bigg(\frac{1}{R_{pvo}}+\frac{1}{R_{bo}}+\widehat{d}_3\Bigg) -\frac{R_{bo}}{R_{pvo}}x_2 \nonumber \\
\dot{\widehat{d}}_3 &=& -\rho_3 x_6 e_6 \label{eq:d3hatdot}
\end{eqnarray}
we get $\dot{W}_6 = - K_6 e_6^2 <0 $ if $K_6>0$. For designing $u_2$, the final Lyapunov function is chosen as follows:
\begin{eqnarray}
W_{4,5} &=& \frac{L_{b}}{2}e_4^2 + \frac{C_{bo}}{2}e_5^2 + \frac{1}{2\rho_2}\widetilde{d}_2^2 \nonumber
\end{eqnarray}
where $e_4=x_4-x_4^d$ and $e_5 = x_5- \alpha_5$
Taking the derivative,
\begin{eqnarray}
\dot{W}_{4,5}= L_b e_4 \dot{x}_4 + C_{bo}e_5(\dot{x}_5 - \dot{\alpha}_5)+ \frac{1}{\rho_2}\widetilde{d}_2(-\dot{\widehat{d}}_2)
\end{eqnarray}
Substituting and rearranging the terms, we get,
\begin{eqnarray}
\dot{W}_{4,5} &=& e_4(\widehat{d}_2-x_5-R_b x_4)+e_5(x_4 + \frac{x_6 - x_5}{R_{bo}}) \nonumber \\ & & - e_5(x_4 u_2 +C_{bo}\dot{\alpha}_5) + \widetilde{d}_2(e_4-\frac{\dot{\widehat{d}_2}}{\rho_2})
\end{eqnarray}
By choosing the following,
\begin{eqnarray}
num_{c2}&=& -e_4(\widehat{d}_2-x_5-R_b x_4)-e_5x_4\nonumber \\ & & - e_5(\frac{x_6 - x_5}{R_bo} -C_{bo}\dot{\alpha}_5) \nonumber \\
den_{c2}&=& e_4x_5- e_5x_4 \nonumber \\
u_2&=&\frac{num_{c2}-K_5e_5^2-K_6e_6^2}{den_{c2}}\label{eq:u2} \\
\dot{\widehat{d}_2}&=&\rho_2e_4 \label{eq:d2hatdot}
\end{eqnarray}
we get, $\dot{W}_{4,5}= -K_4 e_4^2 - K_5 e_5^2 $  which is equivalent to 
$~~~\dot{W}_{4,5}<0$ for $K_4,K_5>0$. 
Thus the total Lyapunov function,
\begin{eqnarray}
\dot{W}= \dot{W}_1 + \dot{W}_{2,3} + \dot{W}_{4,5} + \dot{W}_6 < 0
\end{eqnarray}
Thus, adopting the controllers designed in \eqref{eq:u1}, \eqref{eq:u2} and the estimation update laws in \eqref{eq:d1hatdot}, \eqref{eq:d2hatdot} and \eqref{eq:d3hatdot}, stability of all the states and disturbance observers is guaranteed.

\section{Simulation Results}\label{sec:resultsdo}
In this section, three different cases are investigated related to the change in different external disturbances like temperature, irradiation and load to validate the effectiveness of the proposed disturbance observer based back-stepping control strategy. The DC bus voltage of the SPVDG system is chosen to be 40$V$. The PV system has a peak power rating of 200W at standard temperature and irradiance and the load fluctuates between 145W and 320W. It is also to be noted that in this work, the PV is always operated at maximum power. Table-\ref{table:parametersdo} shows the specifications of the various components used in the SPVDG system along with the controller and observer gains used to implement the proposed technique. It is to be noted that all the simulations were carried out in the MATLAB enviornment.

\begin{figure}[h]
		\centering
		\includegraphics[width=0.8\linewidth]{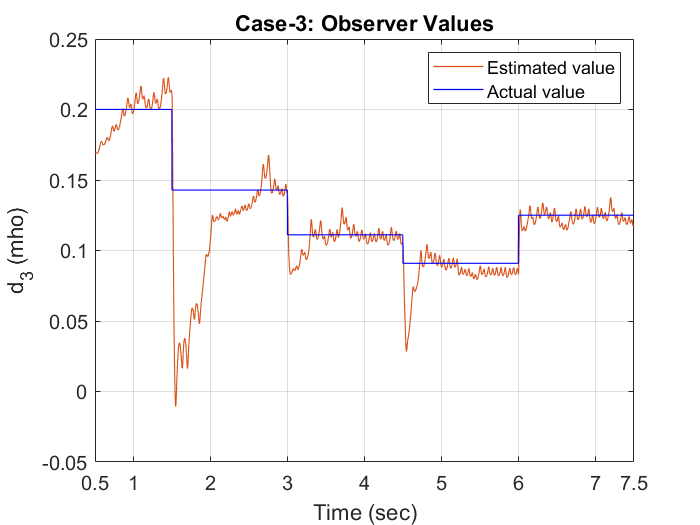}
		\caption{Case-3: Observer values }
		\label{fig:case32}
\end{figure}

 It is clearly visible from figures \ref{fig:case12}, \ref{fig:case22} and \ref{fig:case32} that the disturbance observers start at a random value and update their values in an online fashion leading to simultaneous stabilization of the states by the controllers. The observers continuously monitor the different states and accordingly update their values such that desired control is achieved. 
  
\begin{table}[htbp] 
\caption{Parameter Specification}
\begin{center}
\begin{tabular}{|c|c|}
\hline
\textbf{Subsystem} & \textbf{\textit{Parameter Specification}} \\
\hline
Battery Voltage &  $24 V$ \\
\hline
PV Array at STC & Model: Kyocera Solar KC200GT \\ 
& $V_{OC}=32.9V$, $I_{SC}=8.21A$ \\
& $V_{MPP}=26.3V$, $I_{MPP}=7.61A$ \\  
\hline
DC/DC Converter & $C_{pvi}=3mF$, $C_{pvo}=3mF $, $L_{pv}=10 mH$, \\   & $R_{pv}=0.5\Omega$ , $R_{pvo}=0.1\Omega$ \\
\hline
DC/DC Bidirectional & $C_{bo}=3mF$, $R_{b}=0.5\Omega $, \\
Converter & $L_{b}=10 mH$, $R_{bo}=0.1\Omega $ \\
\hline
Back-stepping & $K_1$=17, $K_2$=0.04, $K_3$=0.04\\
Gains & $K_4$=0.06, $K_5$=0.06, $K_6$=19.0\\
\hline
Observer Gains & $\gamma_1$=10, $\gamma_3$=0.01\\
\hline
\end{tabular}
\label{table:parametersdo}
\end{center}
\end{table}

\subsection{Case-1: Change in Irradiance}
\begin{figure}[h]
	\centering
		\includegraphics[width=0.8\linewidth]{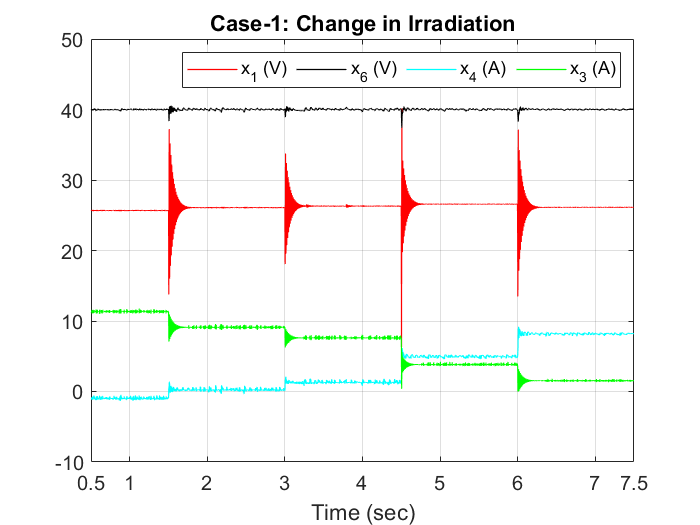}
		\caption{Case-1: Relevant states}
		\label{fig:case11}
\end{figure}

\begin{figure}[h]
		\centering
		\includegraphics[width=0.8\linewidth]{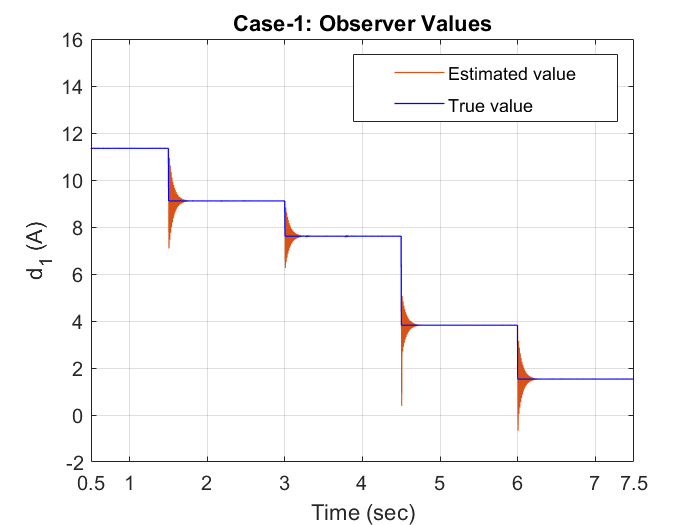}
		\caption{Case-1: Observer values }
		\label{fig:case12}
\end{figure}

In this case, the irradiance of PV panel is changed with time while temperature and load are kept constant at 25$\degree$C and 200$W$. The irradiance changes for every 1.5 second from 1500$W/m^2$ to 1200$W/^2$ and then to 1000$W/m^2$, 500$W/m^2$ and 200$W/m^2$ at 1.5$s$, 3$s$, 4.5$s$ and 6$s$ respectively. Figures \ref{fig:case11} and \ref{fig:case12} show the evolution of pertinent states and estimated disturbance values in this case.

When irradiance changes, both the voltage and current at which maximum power is extracted from the PV panel change. Moreover, the maximum power that can be extracted also reduces with reduction in irradiance. This effect is visible in the PV and battery currents shown in Fig.\ref{fig:case11}. Since load is constant at 200$W$, as the irradiance decreases, the PV inductor current reduces and the battery current increases compensating the reduction in PV current.  During change in irradiance, some transients occur in DC bus voltage but they die out within 100$ms$. Therefore, the DC bus voltage remains constant at 40$V$.

\subsection{Case-2: Change in Temperature}
With increase in temperature, the maximum power extracted from the PV panel reduces, and also the values of MPP current and voltage. In this case, the temperature is varied keeping load and irradiation constant at 266.7$W$, 1000$W/m^2$ respectively. The temperature is varied from 75$\degree$ C to 50$\degree$ C and to 25$\degree$ C, 10$\degree$ C and 0$\degree$ C respectively, at 1.5$s$, 3$s$, 4.5$s$ and 6$s$ respectively.
\begin{figure}[h]
	\centering
		\includegraphics[width=0.8\linewidth]{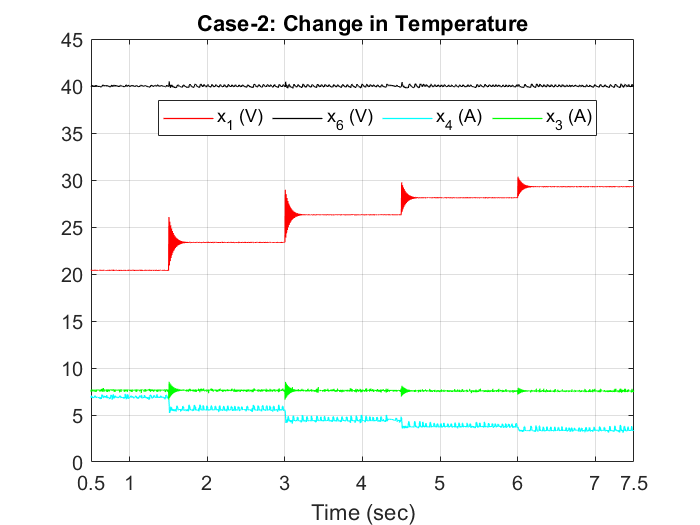}
		\caption{Case-2: Relevant states}
		\label{fig:case21}
\end{figure}

\begin{figure}[h]
		\centering
		\includegraphics[width=0.8\linewidth]{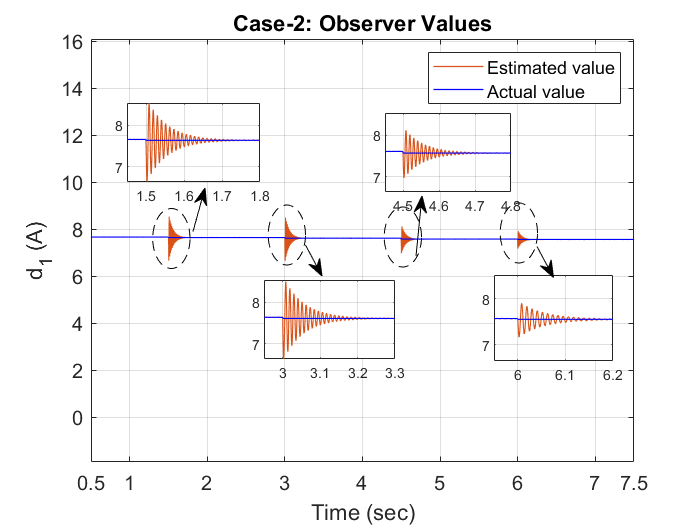}
		\caption{Case-2: Observer values }
		\label{fig:case22}
\end{figure}
Figures \ref{fig:case21} and \ref{fig:case22} show the response of the SPVDG system to change in temperature and the estimation of disturbance $d_1$ respectively. However, the variation in MPP current extracted from PV with change in temperature is less compared to the change in MPP voltage. The grid voltage stays at 40$V$ irrespective of any variation in temperature.  
\newpage
\subsection{Case-3: Change in Load}
In this case, the PV panel characteristics remain constant with constant temperature and irradiation at 1000$W/m^2$ and 25 $\degree$C. However, the load resistance is varied from 5$\Omega$ to 7$\Omega$, 9$\Omega$, 11$\Omega$ and then to 8$\Omega$ at 1.5$s$, 3$s$, 4.5$s$ and 6$s$ respectively.
\begin{figure}[h]
	\centering
		\includegraphics[width=0.8\linewidth]{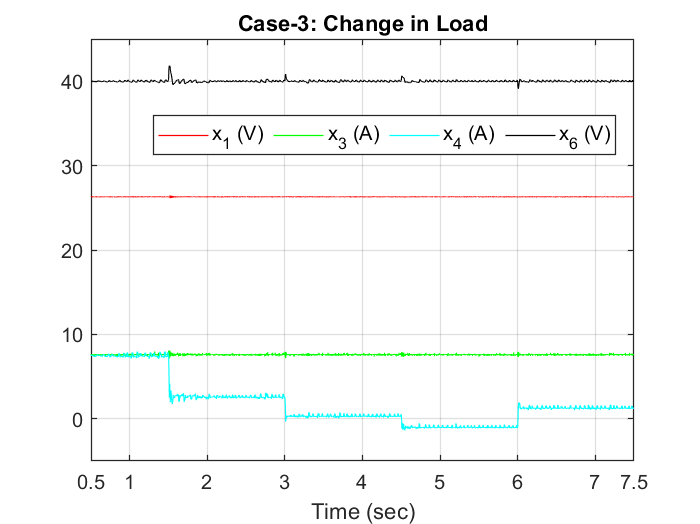}
		\caption{Case-3: Relevant states}
		\label{fig:case31}
\end{figure}

The change in battery current can easily be observed when PV generation is constant and load is changed. Initially, when the load is very high, the battery also supplies positive current to assuage the load requirement and gradually, as the load is reduced the current supplied by the battery reduces and when the load reduces below the maximum power provided by the PV panel, the battery current becomes negative showing that the extra power produced by the PV panel is being used to charge the battery. Its worth noting that both the MPP voltage $x_1$ and grid voltage $x_6$ remain constant. 

In all the cases it can be observed that the grid is hardly perturbed without any deviations from desired value. All the other states and disturbance estimators completely stabilize within 100$ms$. In case-3, it is found that although the disturbance estimation is slower for some variations in load, the DC bus voltage control performance is not affected. Hence, the proposed algorithm is not only fast but also very robust to change in disturbances.

\section{Summary}\label{sec:summarydo}
In this chapter, a disturbance observer based back-stepping control strategy is developed for a standalone PVDG system with battery energy storage. This technique ensures elimination of sensors for measuring PV array output current and load current while preserving all the positive effects of non-linear model based control. The back-stepping controller designed in this chapter results in faster MPP tracking and voltage control. It has been verified for three different cases in presence of intermittencies in irradiance, temperature and load. This controller also ensures appropriate bidirectional power flow depending on the power balance in the standalone system.The simulation results validate the efficacy of the proposed nonlinear control strategy with reduced sensor count resulting in lower system cost while providing better performance. 

%% file: Content/06.tex
\chapter{Conclusion and Future Scope}\label{sec:conclusion}
\section{Conclusion}
In this thesis, we have developed many techniques mainly with a motivation to achieve two objectives :
\begin{itemize}
\item To reduce the number of sensors used in SPVDG system which contributes towards lesser cost of the overall system.
\item To make the SPVDG system more flexible in the sense that it can work in extended ranges of operation while maintaining good performance when subjected to large and sudden disturbances from load and atmospheric conditions.
\end{itemize}

The second chapter discusses a methodology to estimate ambient temperature and irradiation whose data can be used to develop single-step MPPT techniques for PV arrays based on model estimation. When this technique is coupled with a suitable nonlinear control technique, it is possible to achieve MPPT in a record time of 200ms even for drastic changes in atmospheric conditions. In chapter three, we explore a low computation based technique to reduce the sensor requirement in SPVDG systems. We see that when this technique is used, for small changes in irradiation and temperature the MPPT time is around 200ms and voltage control takes around 100ms. However, this method requires the use of linear controllers whose operational range is very low and this technique cannot tolerate heavy disturbances. Hence, in the fourth chapter, we develop a nonlinear back-stepping based controller so that both operational range is enhanced and speed is also maintained. It is seen that it is possible to stabilize the system states around 80ms even in case of large disturbances. But this technique again uses too many sensors and thus, in the fifth, chapter we propose a disturbance observer based strategy to overcome the extreme sensor requirement issue in back-stepping controller.
This technique was seen to stabilize all the system states and observer values around 100ms  which makes it very much viable to implement on SPVDG systems with less sensors and improved performance.

\section{Future Scope}
The work done in this thesis has a lot of potential to be carried forward. Some of the future directions are enumerated as follows:

\begin{itemize}
\item The thesis considers only resistive load   for evaluating the algorithms. All the algorithms can be expanded to more practical loads like constant power load, AC loads.
\item The algorithms and controllers developed in this thesis can be applied to a much bigger    DC system with varied renewable sources like wind, fuel cell and also in the presence of a diesel generator for emergency situations.
\item All the works developed in this thesis consider only an isolated PVDG system. Such issues can be explored also in the grid-connected scenario.
\item As we advance towards more adaptive and disturbance observer based techniques where system information is low, it is difficult to compute the secondary level references. Hence, learning based secondary level reference generation can be explored.
\item The system must be tested using stochastic disturbances after sufficient modeling of the disturbance data.
\end{itemize}

%% file: Content/Publications.tex
\chapter*{\centering List of Publications}
\thispagestyle{empty}
\addcontentsline{toc}{chapter}{List of Publications}

\section*{}
\begin{enumerate}
\item A. Hussain, M. M. Garg, M. P. Korukonda, S. Hasan and L. Behera, "A Parameter Estimation Based MPPT Method for a PV System Using Lyapunov Control Scheme," in IEEE Transactions on Sustainable Energy, vol. 10, no. 4, pp. 2123-2132, Oct. 2019.
\item M. Satapathy, M. P. Korukonda, A. Hussain and L. Behera, "A Direct Perturbation based Sensor-free MPPT with DC Bus Voltage Control for a Standalone DC Microgrid," 2019 IEEE PES Innovative Smart Grid Technologies Europe (ISGT-Europe), Bucharest, Romania, 2019, pp. 1-5.
\item M.P.Korukonda, M.M Garg, A. Hussain and L.Behera, "Disturbance Observer based Controller Design to Reduce Sensor Count in Standalone PVDG Systems", (Submitted to IECON-2020, Singapore)
\end{enumerate}
